\documentclass[useAMS,usenatbib]{mn2e}

\input epsf

\newcommand{\xmm} {{\sl XMM-Newton }}

\newcommand{\rosat} {{\sl ROSAT }}
\newcommand{\asca} {{\sl ASCA }}

\title[A399/401]{\xmm observations of the binary cluster system Abell~399/401}
\author[I. Sakelliou, T.J. Ponman]{Irini Sakelliou,\thanks{E-mail:
irini@star.sr.bham.ac.uk } Trevor J. Ponman\\
School of Physics and Astronomy, University of Birmingham, Edgbaston, Birmingham B15 2TT}

\begin{document}

\pagerange{\pageref{firstpage}--\pageref{lastpage}} \pubyear{2003}
\maketitle

\label{firstpage}
 
\begin{abstract}
Abell~399 and Abell~401 are both rich clusters of galaxies, with
global temperatures of 7.2~keV (Abell~399) and 8.5~keV (Abell~401)
respectively. They lie at a projected separation of $\sim$3~Mpc,
forming a close pair. We have observed the system with the \xmm
satellite. The data of each cluster show significant departures from
our idealised picture of relaxed rich clusters. Neither of the two
contains a cooling flow, and we find that their central regions are
nearly isothermal, with some small-scale inhomogeneities. The image
analysis derives $\beta$-values that are smaller than the canonical
value of 0.65, and the surface brightness distribution is not
symmetric around the central cD galaxies: there are irregularities in
the central $\sim$200~kpc, and asymmetries on larger scales, in that
the intracluster gas in each cluster is more extended towards the
other member of the system. Both clusters host extended radio halos
and a plethora of tailed radio galaxies.  The halo in Abell~399
appears to be correlated with a sharp edge apparent in the \xmm
images, and a region of harder X-ray emission. There is also evidence
for enhanced X-ray flux in the region between the two clusters, where
the temperature is higher than our expectations.

Although tidal or compression effects might affect the large scale
structure of the two clusters, we show that these cannot account for
the distortions seen in the inner regions. We argue that the
reasonably relaxed morphology of the clusters, and the absence of
major temperature anomalies, argues against models in which the two
have already experienced a close encounter. The properties of the
intermediate region suggests that they are at an early stage of
merging, and are currently interacting mildly, because their
separation is still too large for more dramatic effects. The
substructure we find in their inner regions seems to point to their
individual merging histories. It seems likely that in the
Abell~399/401 system, we are witnessing two merger remnants, just
before they merge together to form a single rich cluster of
galaxies. This picture is consistent with recent numerical
simulations of cluster formation.

\end{abstract}


\section{Introduction}

It is now common wisdom that clusters of galaxies are formed
hierarchically, by the merging of smaller mass units, preferentially
along large-scale filaments. During such violent events one expects
physical processes to take place that have significant impact on the
properties of the constituents of clusters (i.e., gas, galaxies), that
would define their subsequent evolution, as well as the energy and
entropy budget in the largest structures in the Universe. During
mergers, shock waves propagate within the intracluster medium (ICM),
thus increasing its thermal energy and entropy. The cluster galaxies
may be severely stripped, leaving their metal rich interstellar media
(ISM) behind in the cluster, contributing in this way to the
enrichment of the ICM (e.g., Acreman et al. 2003). Cooling flows may
get displaced or destroyed entirely (e.g., G\'{o}mez et al. 2002) .

\begin{figure}
\begin{center} 
\leavevmode 
\epsfxsize 1.05\hsize
\epsffile{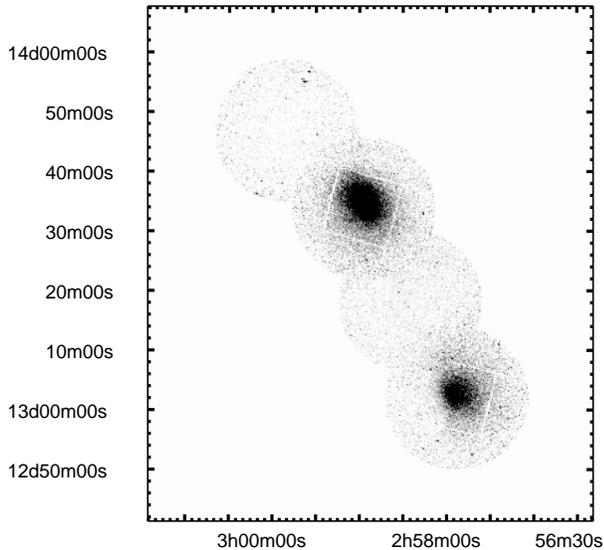}
\caption{A mosaic image of all 4 pointings in the (0.5-10)~keV energy
range. Background subtracted images from the two MOS instruments (MOS1
and MOS2) are superimposed. The pixel size is 8~arcsec.  Abell~401 is
to the North-East of the image.}\label{mosaic}
\end{center} 
\end{figure}

Numerical simulations (Schindler \& M\"{u}ller 1993, Roettiger, Burns,
\& Loken 1996; Ricker 1998; Takizawa 1999; Ritchie \& Thomas 2002;
Burns et al. 2004)
have advanced to such an extent that they can provide us with detailed
temperature maps and synthetic X-ray images of clusters under
formation, at each stage of their evolution. All this work has
demonstrated that cluster mergers result in dramatic substructure,
visible in the X-ray images and spectra of clusters of
galaxies. Initially, during the first approach of two clusters a
compression and/or shock wave is developed in the interface between
them. When the two cores collide, the temperature and luminosity of
the system reaches its maximum. As the two gravitational potentials
settle down to form a single one, two lens-shaped shock waves
propagate outwards, along the direction of merging. Soon afterwards, a
single remnant is formed.

Unfortunately, we have not been able yet to test observationally in a
systematic manner, that cluster mergers proceed as theory
predicts. Markevitch et al. (1998) set the framework for such
investigations, and revealed the gross spectral characteristics of
disrupted clusters. Additionally, they were able to comment on the
statistics of mergers, but their data were lacking the necessary
quality to probe the details of the observed structures. But the
results of some recent observational work have been encouraging (e.g.,
Sun et al. 2002; Kempner, Sarazin, \& Ricker 2002), as a hotter bar,
for example, has been found in a few cases separating the two
colliding units, providing the means to calculate the relative
velocity of the merging subunits. A detailed comparison of a sample
with the numerical simulations is still lacking, and it is unclear
whether the simulations include all the necessary physics and are able
to model the problem correctly. There are a few examples of cluster
properties that have been discovered by the observations, and had not
been predicted by the simulations.  Recently, for example, X-ray
observations have uncovered features (cold fonts; e.g., Markevitch et
al. 2000), that have been now explained as signatures of merging
clusters. Cold fronts had not been predicted by the simulations,
mainly due to the lack of the adequate resolution and the mishandling
of the cooling functions. Only recently some numerical simulations
have managed to reproduce them and associate their presence to the
motion of a galaxy or group through the ICM of another cluster, and
explain the observations (Bialek, Evrard, \& Mohr 2002).

Ideally, one would like to construct a sample of clusters that are at
different stages of their evolution, and trace the merger
event. Unfortunately, finding clusters at different stages of the
merging sequence is not trivial. Projection effects or past inaccurate
determinations of their physical properties influence severely their
classification and taxonomy in one of the stages. Abell~399 and
Abell~401 appear at a first sight as a very good example of two
clusters at early stages of merging. They are equally rich, and past
X-ray missions have found them to be both at temperatures between 7
and 8~keV. They have been the subject of past investigations and have
been observed in different wavelengths, from the optical to X-ray.

Their projected separation is
$\sim$36~arcmin~$\sim$~2.96~Mpc\footnote{Throughout this paper we use
$H_0=71~{\rm km \ s^{-1} \ Mpc^{-1}}$, $\Omega_{\rm M}$=0.3, and
$\Omega_{\rm \Lambda}$=0.7.}, approximately 1-2 times their virial
radii.  The line-of-sight velocity difference of the two is of the
order of $\sim 700 \; {\rm km \ s^{-2}}$ (Girardi et al. 1997). The
velocity dispersions of Abell~399 and 401 are $\sim 1180$ and $\sim
1110$~${\rm km \ s^{-2}}$ (Oegerle \& Hill 2001) respectively.
Dynamical models of the system based on the dynamics of the galaxies
around the Abell~399/401 system reached the conclusion that it
consists a bound cluster pair (Oegerle \& Hill 1994), and that the
galaxies of both clusters should move in the total gravitational
potential of both clusters. Additionally, since their line-of-sight
velocity difference is smaller than their velocity dispersions, it is
probable that the encounter between the two clusters is taking place
essentially on the plane of the sky.  Basic properties of the clusters
are given in Table~1.

\begin{table}
\begin{center}
\caption{Target Information}\label{target_info}
 \begin{tabular}{ccccccccc} \hline \hline

Cluster		&
z		&
kpc/''		&
$D_L$ (Mpc)
\\

\hline

A399		&
0.0724		&
1.36 		&
322.4
\\

A401		&
0.0737		&
1.38		&
328.5
\\

\hline

\end{tabular}
\end{center}
\end{table}

\begin{table*}
\caption{Pointing Information}\label{obs_info}
\begin{center}
 \begin{tabular}{ccccccccc}   \hline \hline

(I)		&
(II)		&
(III)		&
(IV)		&
(V)		&
(VI)		&
(VII)
\\

Rev 		&
Obs		&
$\alpha$~(2000) 	&
$\delta$~(2000) 	&
Instr.			&
$Exp$		&
$Exp_{corr}$		&

\\
		&
		&
deg 		&
deg 		&
		&
ksec 		&
ksec		&

\\
\hline

0127		&
0112260101	&
44.4491255188460	&
13.0518334856848	&
MOS1			&
14.166 	&
11.429 	&
\\

	&
	&
	&
	&
MOS2	&	
14.170	&
11.599	&
\\

	&
	&
	&
	&
PN	&	
9.188	&
5.249 	&
\\

0127		&
0112260201	&
44.5827088537386		&
13.3186112665767		&
MOS1				&
18.231		&
18.133 		&
\\

	&
	&
	&
	&
MOS2	&	
18.234	&
18.152	&
\\

	&
	&
	&
	&
PN	&	
12.505 	&
12.395 	&
\\

0395		&
0112260301&
44.7636255225170	&
13.5472223803563    &
MOS1			&
12.953	&
12.948	&
\\

	&
	&
	&
	&
MOS2	&	
12.959 	&
12.897	&
\\

	&
	&
	&
	&
PN	&	
8.086 	&
8.063	&
\\

0395		&
0112260401	&
44.9844588584281	&
13.7638057162178	&
MOS1			&
11.913 	&
11.882	&
\\

	&
	&
	&
	&
MOS2	&	
11.917	&
11.829	&
\\

	&
	&
	&
	&
PN	&	
7.143	&
7.143 	&
\\

\hline

\end{tabular}
\vspace{0.2cm}
\begin{minipage}{16cm}
\small NOTES-- (I): revolution number; (II): observation number; (III):
pointing $Right Ascension$ ($\alpha$) in degrees; (IV) pointing
$Declination$ ($\delta$) in degrees; (V) EPIC Instrument; (VI) Exposure time
(live time for the central CCD); (VII) Reduced Exposure time, after
the subtraction of the bright background flares (see text for more
details).

\end{minipage}
\end{center}
\end{table*}

\begin{figure*}
\begin{center}
\setlength{\unitlength}{1cm}
\begin{picture}(8,7)
\put(-6.5,7.5){\includegraphics{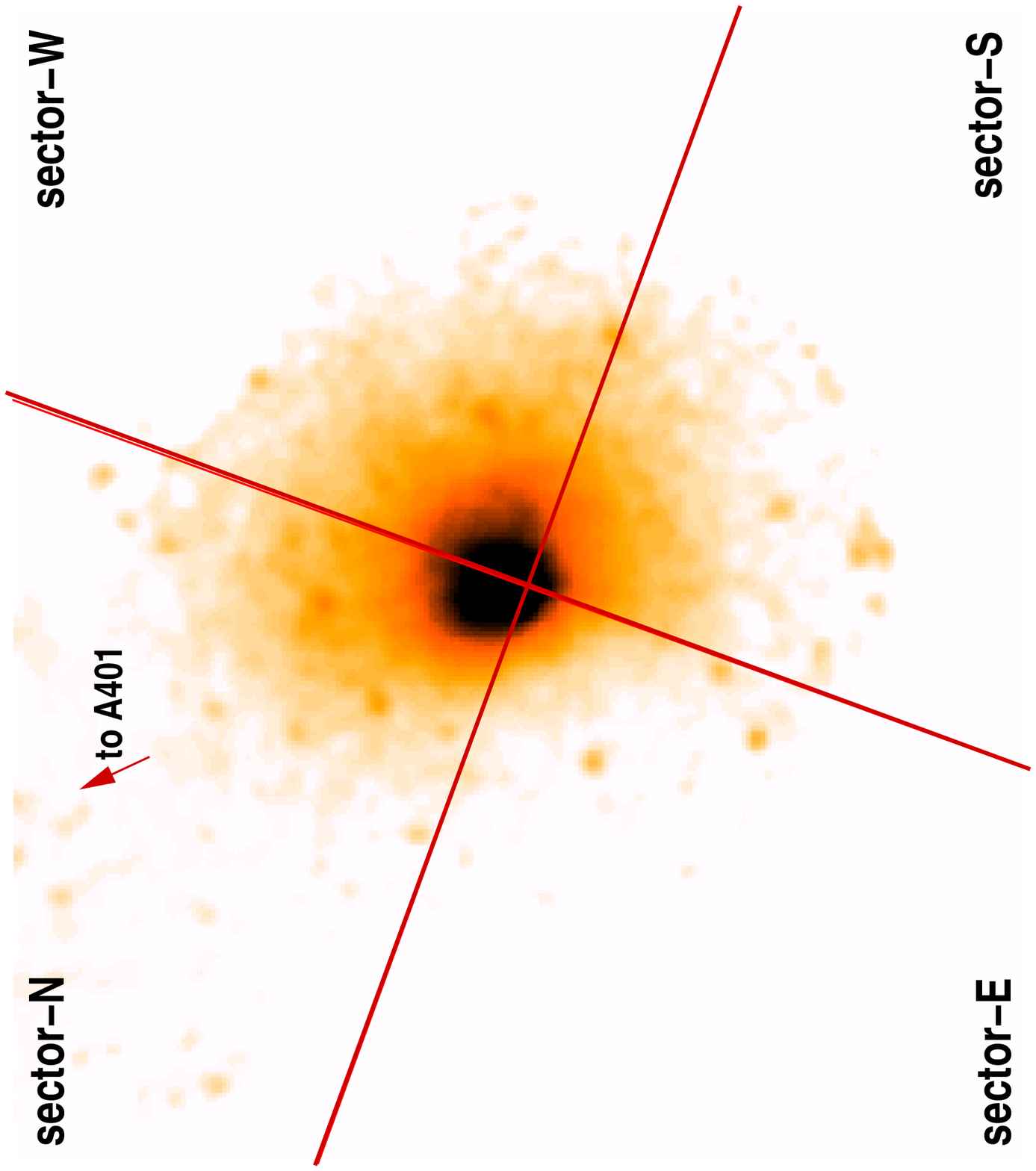}}
\put(3,7.5){\includegraphics{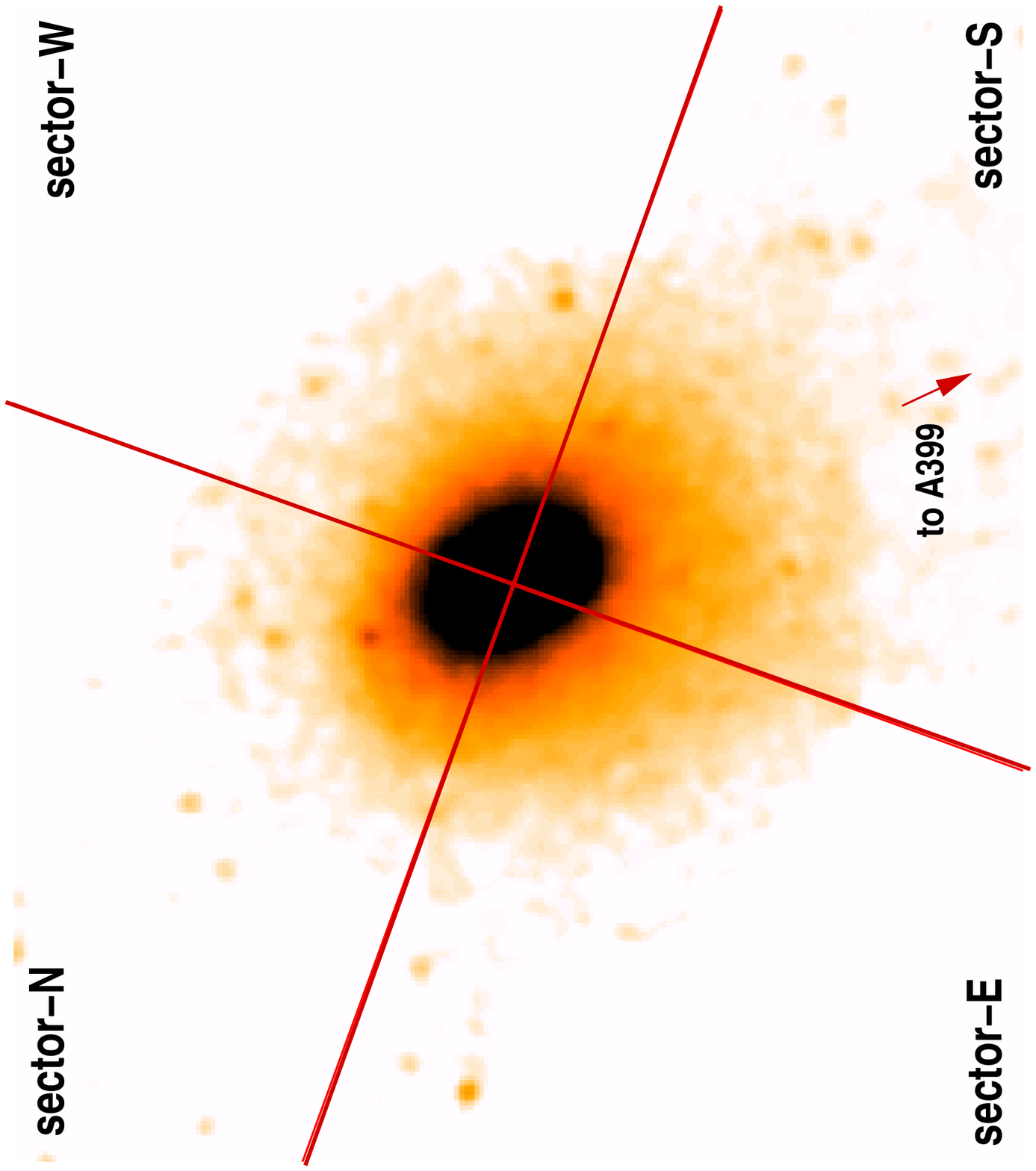}}
\end{picture}
\end{center}
\caption{\xmm images of Abell~399 (left panel) and Abell~401 (right
panel). The images have been smoothed with Gaussian kernels of
$\sigma$ = 2~pixels=16~arcsec. The four sectors, centred on the
central cDs in each cluster, and used in the subsequent analysis, are
shown.}\label{A399pies}
\end{figure*}

Previous studies have reached contradictory conclusions about the past
history of the pair. The first attempts to reveal signs of
interactions in the region between the two with the {\it Einstein}
observatory did not lead to any strong conclusions (Ulmer \& Cruddace
1981).  The \asca satellite found evidence for some enhancement in
the X-ray flux above what is expected from just the superposition of
the two clusters (Fujita et al. 1996). A slight temperature increase
in-between them was also recorded. Both results led Fujita et
al. (1996) to suggest that Abell~399/401 is a pre-merging
pair. However, this scenario provides no explanation for the lack of
cooling flows in the cores of both clusters. X-ray observations (e.g.,
Peres et al. 1998) have found that their centres do not host cooling
flows as the mass accretion rate is zero. Additionally, the model
of Fujita et al. does not tie in well with the past \rosat
observational facts that both clusters appear disrupted : the work 
of Slezak, Durret, \& Gerbal (1994), for example, found evidence
for `substructure' in Abell~401, and the \rosat HRI detector (Fabian,
Peres, \& White 1997) revealed a linear structure that emanates from
the centre of Abell~399 and points towards Abell~401. This feature
lead to the suggestion that the two clusters have already encountered
each-other, and are now moving apart.

In order to recover the dynamical state of this system, decide on its
past history and future evolution, and derive vital information that
would help us to test the results of the numerical simulations of
merging clusters, we observed Abell~399/401 with \xmm. The
observations are presented in Section~2. Section 3 is devoted to the
presentation of the properties of each cluster individually, as found
from the \xmm data and analysis. In Section~4 the X-ray properties of the
region between the two is investigated, while in Section~5 the radio
properties of the clusters are presented, and their large scale
environment is discussed in Section~6. Finally, in the discussion section
(Section~6) we present models for the dynamical states of both clusters that
can fit well the \xmm radio and optical results.

\section{\xmm Observations}

The Abell~399/401 system was observed by the \xmm observatory in 4
pointings, covering a large area around it. Information on the
different observations is given in Table~\ref{obs_info}.  Observation
0112260101 (101 hereafter) was centered on Abell~399, 0112260301 (301)
on Abell~401, 0112260201 (201) in between the two clusters, and
0112260401 (401) to the North of Abell~401. During each observation
the MOS instruments were operating in the PrimeFullWindow mode, and
the PN in the PrimeFullWindowExtended. The thin filter was used for
the 101, 201, and 401 observations, and the medium filter was used
during the 301 observation. A mosaic of all the pointings is presented
in Fig.~\ref{mosaic}.

\subsection{Data Reduction}

All data sets were processed with the \xmm SAS v5.3. {\sc emchain} and
{\sc epchain} were used to obtain the calibrated event lists for the
MOS and PN instruments respectively. During the processing, a search
for new bad pixels was allowed by switching on the {\sc withbadpixfind}
parameter. The calibrated event files were subsequently filtered to
keep only the events with {\sc pattern=0}. Additionally, we clean the
event lists for periods of high background. This cleaning process
reduced the exposure times to those presented in column VII of
Table~\ref{obs_info}.

\subsection{Background Treatment}

Background images and spectra were generated from the `blank-sky'
event lists (D. Lumb's background files; Lumb 2002). The coordinate
frames of these fields were converted to the corresponding frames for
each pointing. The background event lists were filtered for {\sc
pattern} in the same way as the data, and periods of high background
levels that are still present in D. Lumb's files, were removed by
applying a 3-$\sigma$ cut-off. Subsequently, the background events
were scaled to match the background levels of each instrument and
observation by scaling the out-of-field events as in Pratt et
al. (2001). The scaling factors we use for MOS1, MOS2 and PN are:
1.00, 0.95, 1.15, for the 101 observation; 1.06, 0.95, 1.15 for 201;
1.10, 1.03, 1.44 for 301; 1.13, 1,05, 1.30 for the 401 one. These
scaling factors are consistent with the factors found for other
observations.

\subsection{Image Mosaic}\label{s_mosaic}

Figure~\ref{mosaic} shows the mosaic of all four pointings in the
(0.5-10)~keV energy band. For this mosaic we use the exposure
corrected, background subtracted images, that we generate and use in
the subsequent analysis. The production procedure we follow will be explaned
later in this paper (Section~3.1). For the image mosaic of Fig.~\ref{mosaic}
we use only the images from the two MOS instruments (MOS1 and MOS2).
The mosaic was made using the SAS task {\sc emosaic}, without applying
an exposure correction.  

As seen in Fig.~\ref{mosaic} the pointings were chosen so that they
lie along the line that connects the two cluster cores, and each
cluster was at the centre of one observation.

\section{Average Cluster Properties}

In this section we derive the properties of each cluster individually,
and compare them with past results from other X-ray satellites of the
same and similar clusters.

\subsection{Spatial Analysis}

\begin{table}\label{tab_spatial}
\begin{center}
\caption{Spatial Analysis Results}
\begin{tabular}{cccc} \hline\hline

Cluster 		&
Sector		&
$r_{\rm c}$	&
$\beta$	\\

		&
		&
(arcmin)	&
		\\

\hline

A399		&
global	&
1.903$\pm$0.007		&
0.498$\pm$0.001		\\

A401		&
global	&
2.048$\pm$0.004		&
0.590$\pm$0.001		\\

\hline

A399		&
N		&
2.113		&
0.512		\\

		&
W		&
3.129		&
0.574		\\

		&
S		&
2.143		&
0.497		\\

		&
E		&
0.997		&
0.510		\\

A401		&
N		&
2.716		&
0.727		\\

		&
W		&
2.343		&
0.649		\\

		&
S		&
1.937		&
0.545		\\

		&
E		&
1.479		&
0.515		\\

\hline

\end{tabular}
\end{center}
\end{table}

In order to obtain the overall spatial characteristics of both
clusters, and compare them with previous findings, we initially assume
spherical symmetry and fit the surface brightness distributions with
the traditional $\beta$-model (e.g., Cavaliere \& Fusco-Femiano 1976).

For the purpose of any subsequent spatial analysis, we first create
background subtracted and exposure corrected images of both clusters
in the (0.5-10.0)~keV energy range. The background and exposure
correction is performed as follows. Images and exposure maps are
generated from the clean and filtered event lists in narrow energy
bands from 0.5 to 10.0~keV [`narrow-energy-range' (NER) images]. Each
energy band is 0.5~keV wide. Similar NER background images are created
from the blank-sky event files. Subsequently, each NER background
image is smoothed and scaled, using the appropriate scaling factors
for each observation found in Section~2.2. The smoothed and scaled NER
background image is then subtracted from the data image. Each
instrument is treated independently.  Then, each NER background
subtracted image is corrected for exposure and vignetting, by dividing
with the appropriate exposure map. Finally, all the background
subtracted, vignetting and exposure corrected NER images are co-added
to create one image for each instrument in the (0.5-10.0)~keV energy
range.

We first use the above created images to derive the global spatial
properties of each cluster in Section~3.1.1, and later (Section~3.1.2) to
investigate their azimuthal variations.

\subsubsection{Global Properties}

We construct surface brightness profiles for both clusters, by
accumulating counts in 40 concentric circular annuli of width
15~arcsec, around each cluster centre. These profiles are centred on
the peak of the X-ray emission, which coincides with the location of
the large cD galaxies, that reside within the core regions of both
clusters. These locations are marked on Fig.~2. We include in
the profiles only counts that come from regions of the CCDs that were
exposed for more than 0.5~percent of the total exposure time. The
profiles are fit in {\sc sherpa} by the $\beta$-model. For each
cluster, the accumulated profiles from each instrument are fitted
simultaneously, leaving each normalisation as a free parameter.  On
the other hand, the core radius ($r_{\rm c}$) and the
$\beta$-parameter are linked.

\begin{figure*}
\begin{center} 
\leavevmode 
\epsfxsize 0.45\hsize
\epsffile{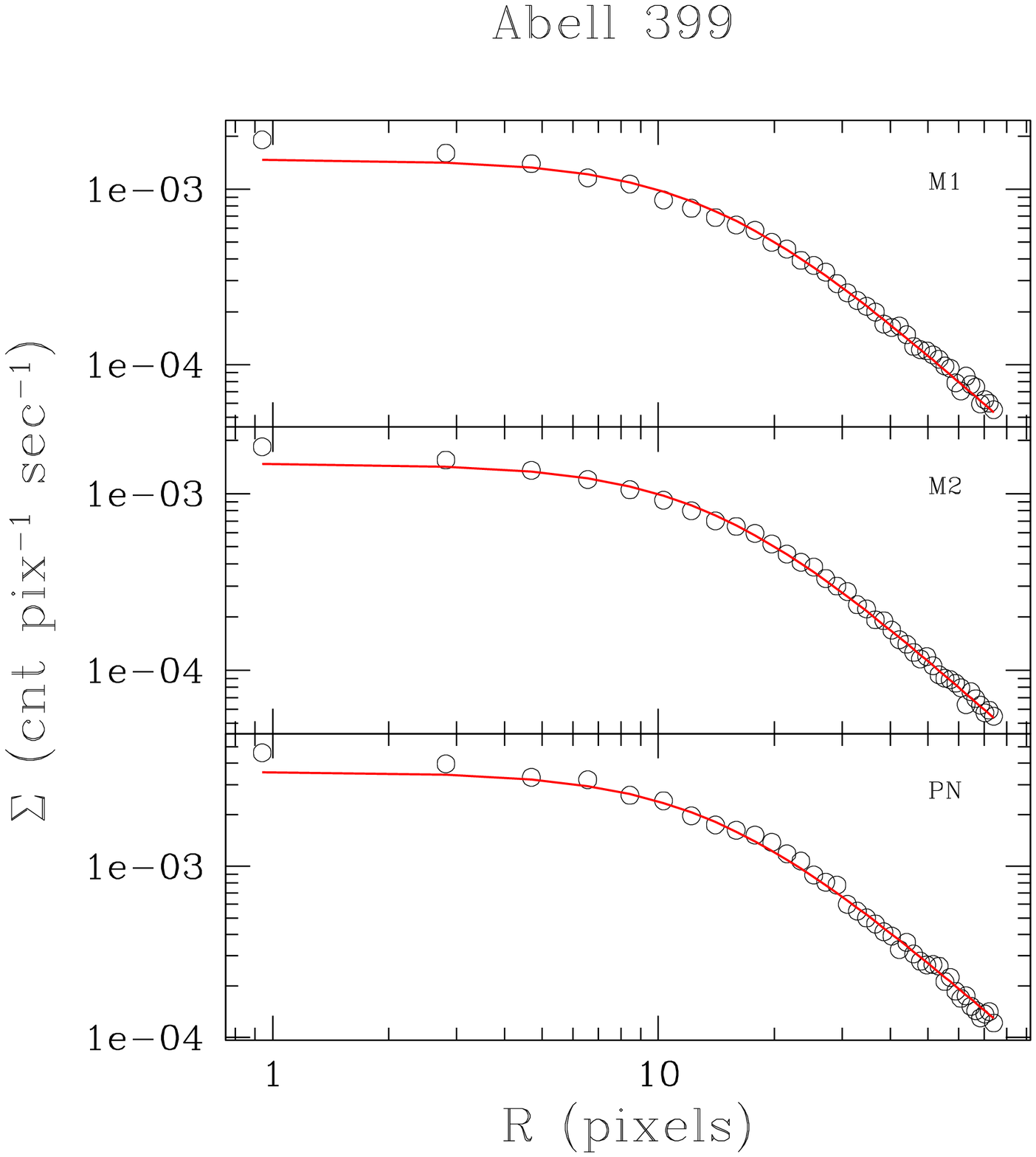}
\leavevmode 
\epsfxsize 0.45\hsize
\epsffile{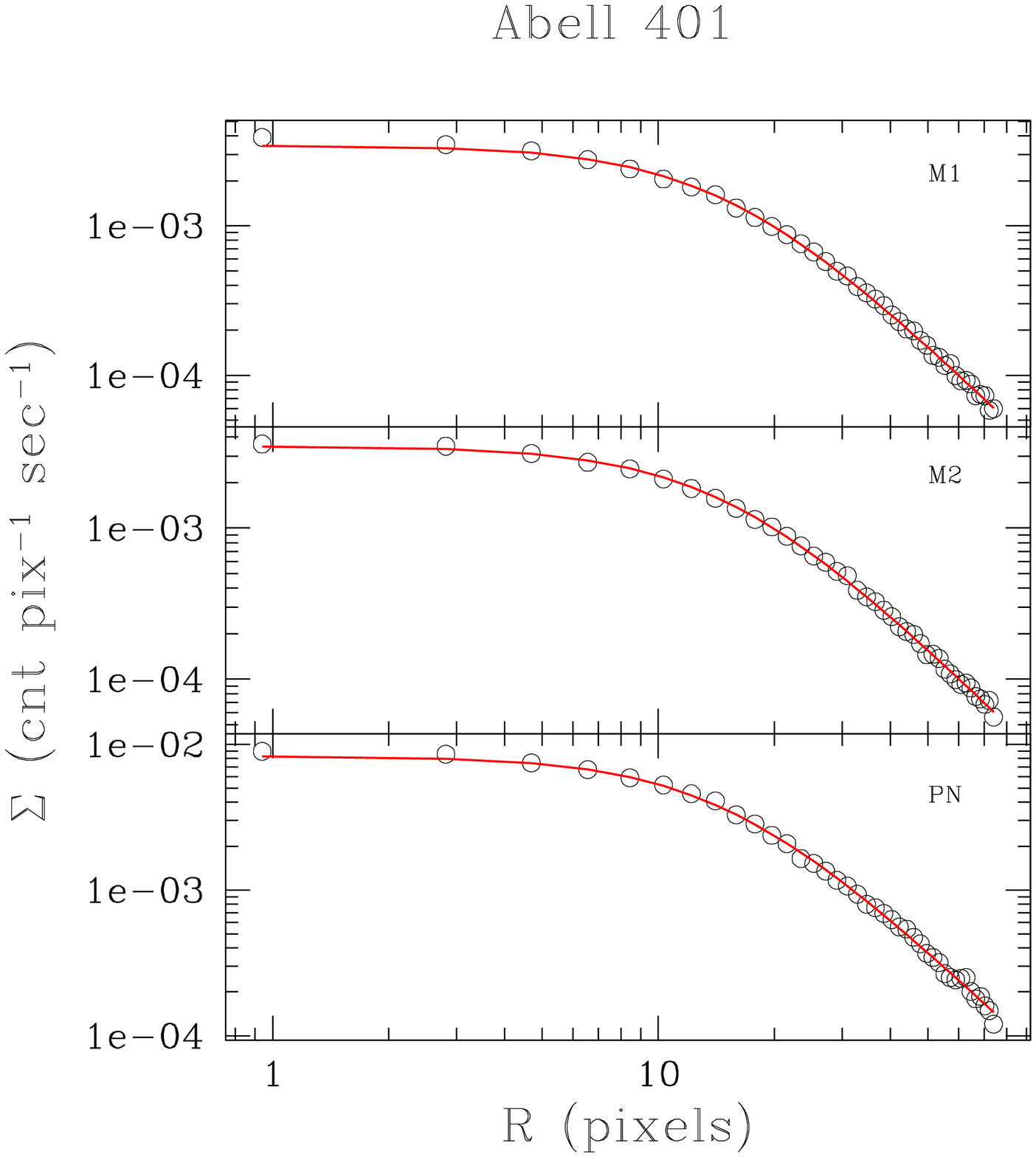}
\end{center}
\caption{Surface brightness profiles of Abell~399 (left panel) and
Abell~401 (right panel). The accumulated profiles from each EPIC
instrument are shown. The best fit $\beta$ model is also
plotted individually. The pixel size is 8~arcsec. }\label{profiles}
\end{figure*}

The fits with a single $\beta$-model give a core radius and a beta
value of $r_{\rm c}= 1.903$~arcmin = $155.3$~kpc, and $\beta$=$0.498$
with $\chi^{2}$/d.o.f = 383/115 for Abell~399, and $r_{\rm c}$=
$2.048$~arcmin = $169.6$~kpc, $\beta$=$0.590$ ,
$\chi^{2}$/d.o.f=343/115 for Abell~401.

Figure~\ref{profiles} shows the azimuthal average surface brightness
profiles and the best fit single $\beta$-model. The data and model for
each instrument are shown independently. It is clear from this figure
that a single $\beta$-model does not represent well the inner regions
of Abell~399, where an excess above the model is apparent. The
addition of a Gaussian to describe the central excess improves the
fit, although it is still a poor description of the data, providing a
new statistic of $\chi^{2}$/d.o.f=300/111. During the last fitting
procedure, the {\it full-width-half-maximum} ($FWHM$) of the Gaussian,
$r_{\rm c}$ and $\beta$ of the $\beta$-model, are left free to be
determined by the fit. Such a composite fit gives a low $\beta$ value,
similar to the single component $\beta$-model fit, with
$\beta$=$0.516$ and a similar core radius of $r_{\rm c}$=
$2.176$~arcmin = $177.6$~kpc. We find that the best fit Gaussian has a
width of $FWHM$= $1.147$~arcmin = $93.6$~kpc.

We find similar behaviour in Abell~401, where a smaller central excess
is also seen. If we add a Gaussian, the properties of the best fit
$\beta$-model do not change significantly; the new values are $r_{\rm
c}$ = $2.116$~arcmin = $175.2$~kpc, and $\beta$=$0.596$, with
$\chi^{2}$/d.o.f=311/111.  For the central Gaussian the fit gives:
$FWHM$= $0.907$~arcmin = $75.1$~kpc.

As can be seen in Table~3, the modelling of the surface brightness
distributions of both clusters results in small $\beta$ values
compared to the canonical value of 0.65 that describes on average the
profiles of rich clusters. These values do not change significantly
either with the addition of the Gaussian as stated above, 
nor with the exclusion of the inner (15-20)~arcsec from the fit.

Neither of the two central galaxies is known to host an active
nucleus, whose presence could be invoked to explain the requirement
for an extra, but small, central component.  It is also apparent that
the results of the composite fits support this statement: we find that
the central emission is not due to a point source, as the width of the
best-fit Gaussians is larger than the \xmm {\it point-spread-function}
(PSF). Another plausible explanation for the observed central excess
could be that the it is due to emission from dense and cool gas that
resides in the cluster cores. However, the excesses are not as large
as those encountered in massive cooling flow clusters. We will come
back to this point in Section~3.2, where we investigate spectroscopically
the possibility of a cooling flow.

\begin{figure*}
\begin{center} 
\leavevmode 
\epsfxsize 0.45\hsize
\epsffile{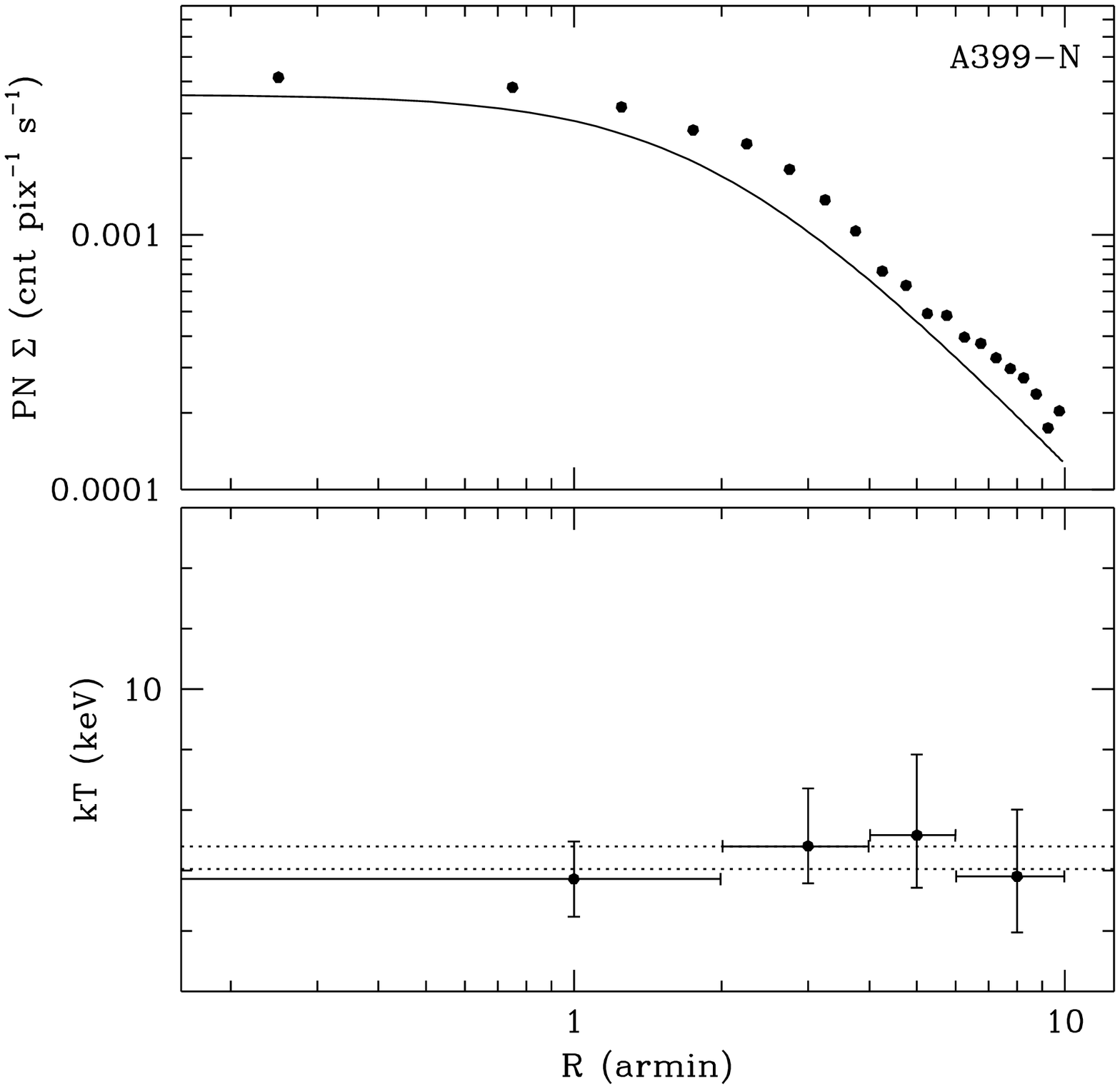}
\leavevmode 
\epsfxsize 0.45\hsize
\epsffile{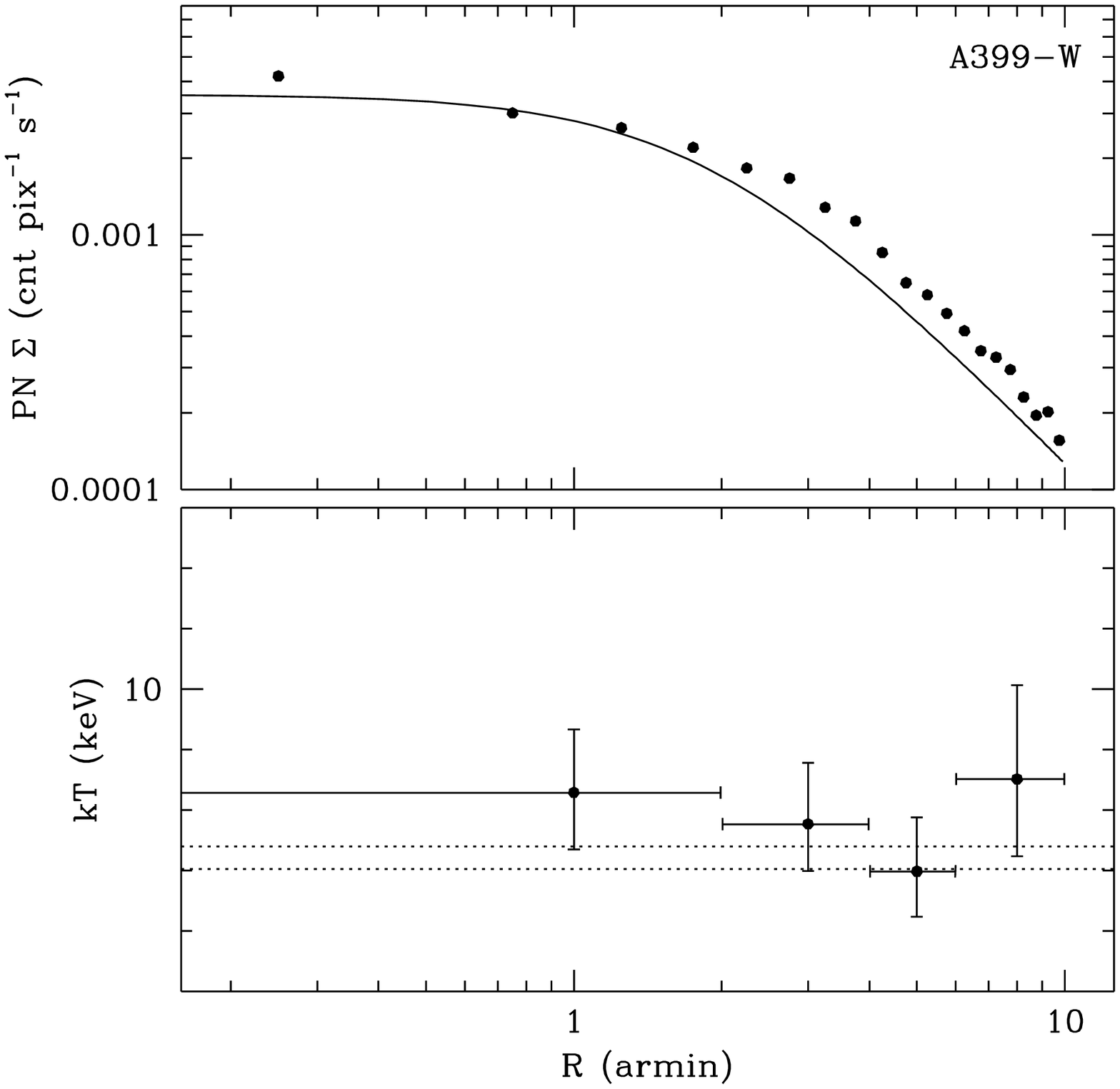}
\end{center}
\begin{center} 
\leavevmode 
\epsfxsize 0.45\hsize
\epsffile{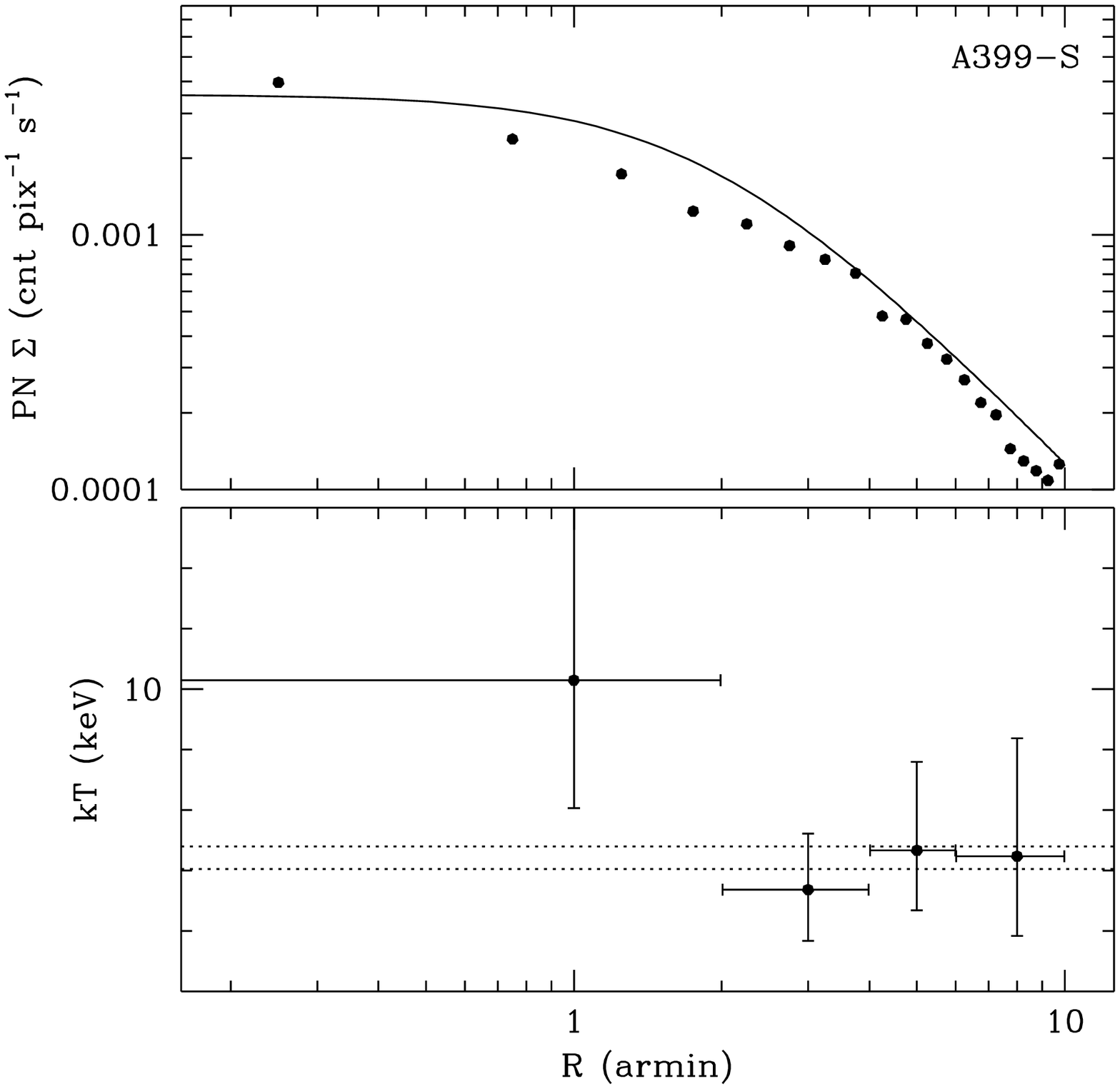}
\leavevmode 
\epsfxsize 0.45\hsize
\epsffile{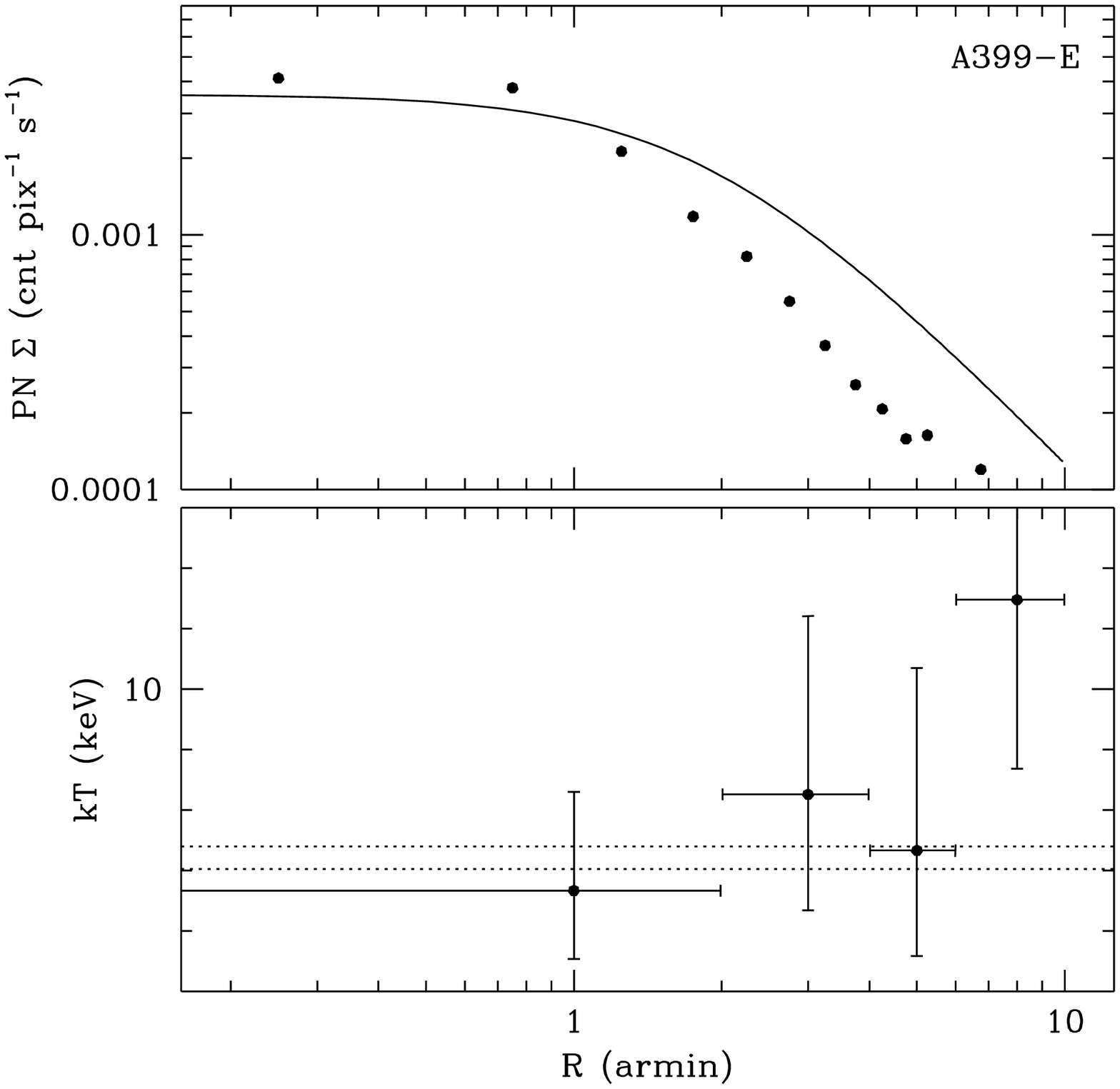}
\end{center}
\caption{Surface brightness ($\Sigma$) and temperature ($kT$) profiles
for each of the four sectors in Abell~399. The surface brightness as
registered on the PN detector only are shown. The size of the errors
is similar to the size of the data points. The solid line in the
surface brightness plots represents the global $\beta$-model found in
Section~3.1.1. The errors shown for the temperature are the 90~percent
errors.}\label{A399T_prof_pies}
\end{figure*}

\begin{figure*}
\begin{center} 
\leavevmode 
\epsfxsize 0.45\hsize
\epsffile{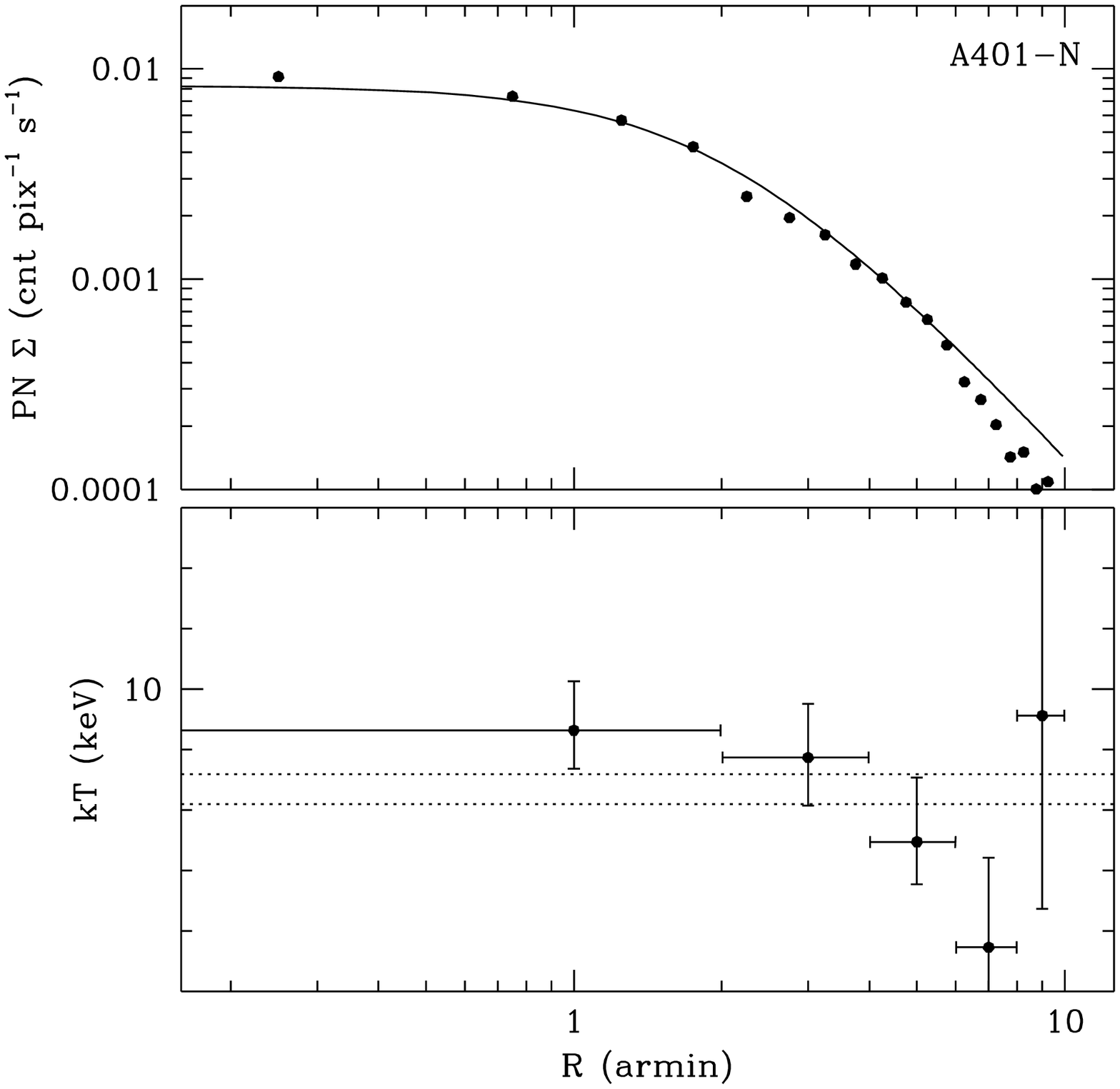}
\leavevmode 
\epsfxsize 0.45\hsize
\epsffile{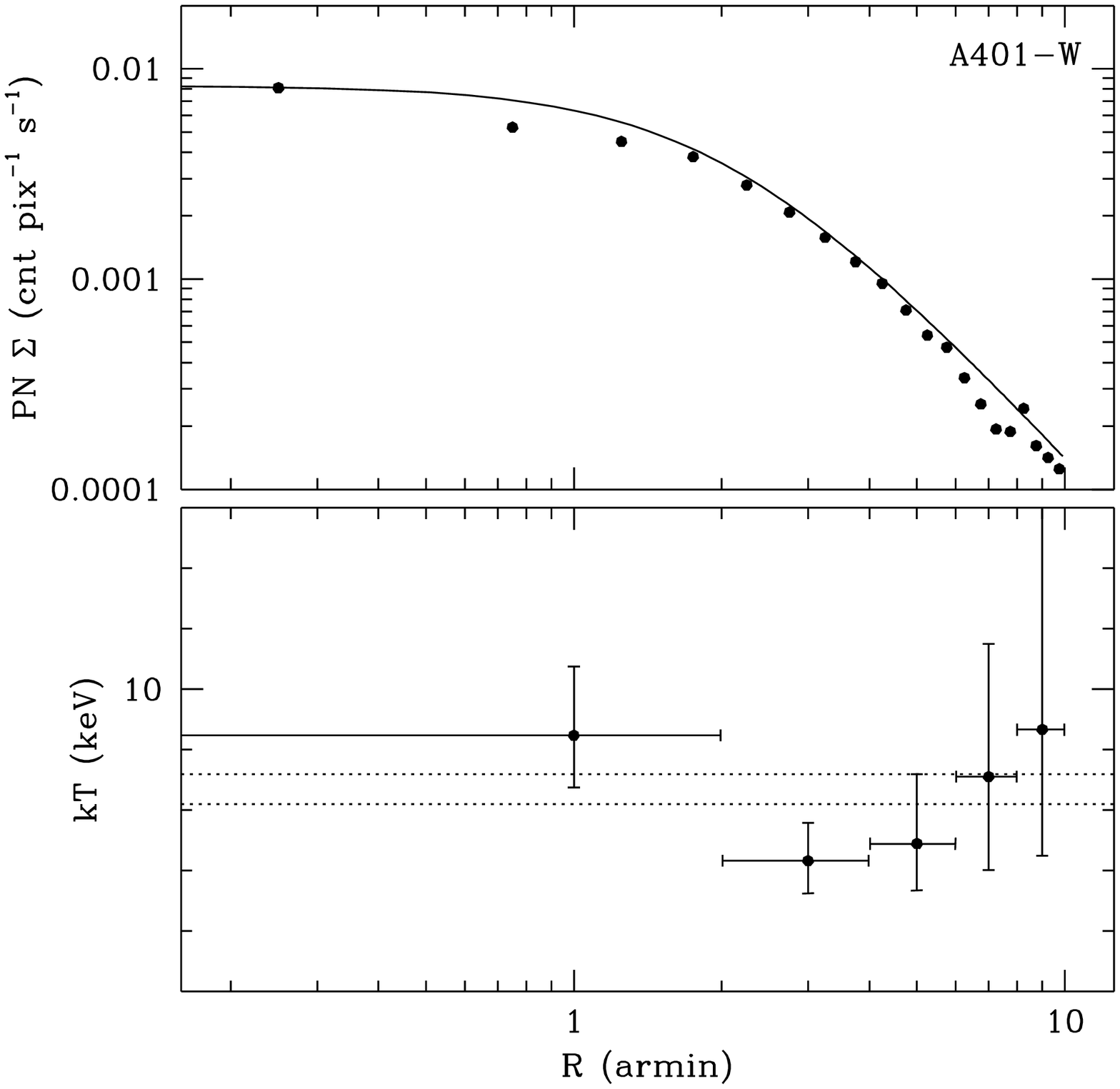}
\end{center}
\begin{center} 
\leavevmode 
\epsfxsize 0.45\hsize
\epsffile{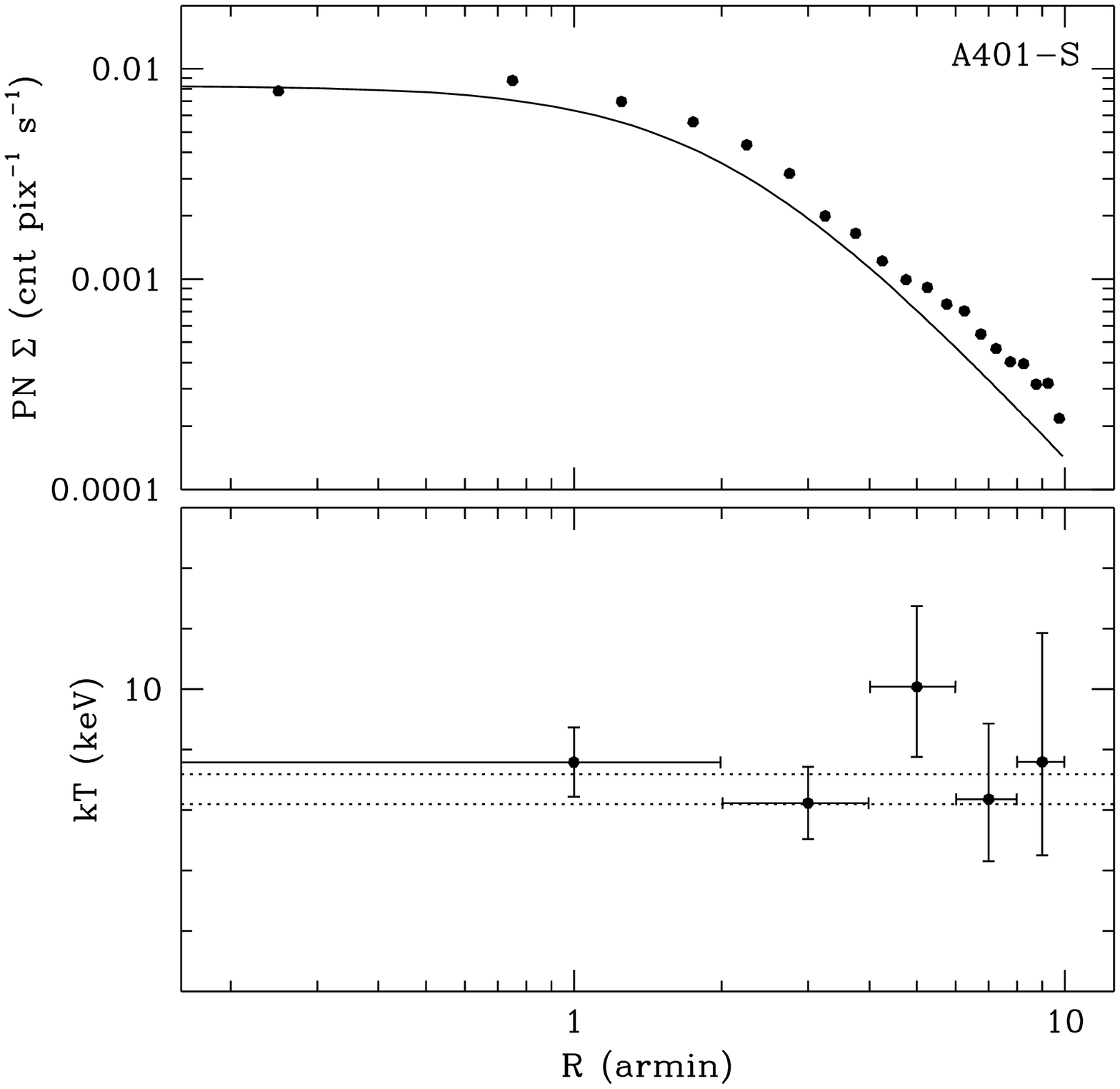}
\leavevmode 
\epsfxsize 0.45\hsize
\epsffile{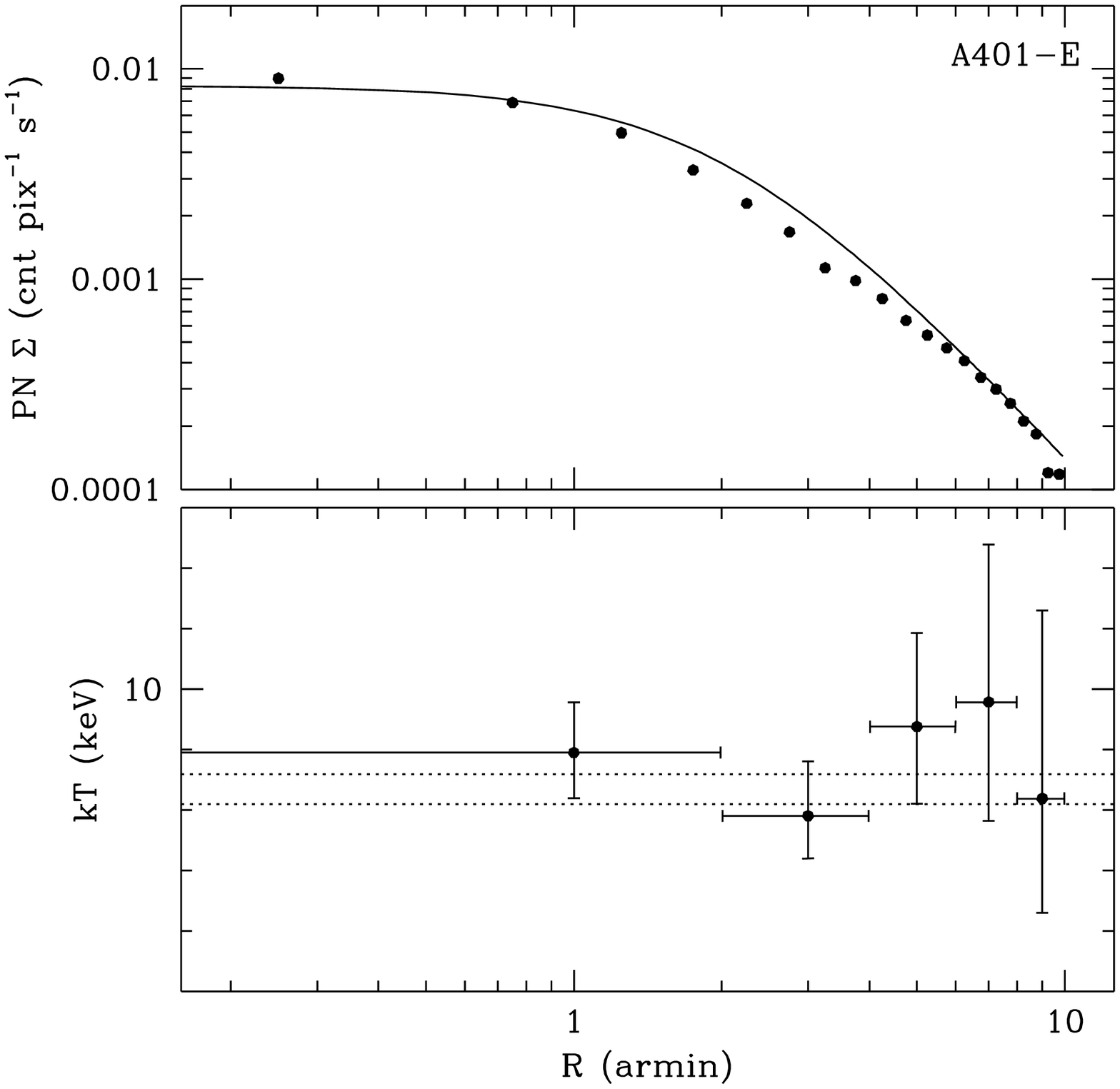}
\end{center}
\caption{As in Fig.~\ref{A399T_prof_pies} but for Abell~401. }\label{A401T_prof_pies}
\end{figure*}

The global fits for both clusters result in large values of the
$\chi_{\nu}$ ($\chi_{\nu} \sim 3$). Under these circumstances
the use the normal $\Delta\chi^2=1$ criterion to define parameter
errors (Lampton, Margon \& Bowyer 1976)
is not valid, and naive application of this prescription
is likely to yield unrealistically small confidence regions.
Although there is a systematic excess above the model in the cores
of both clusters (Figure~\ref{profiles}), most of
the excess scatter about the fitted model which gives rise to
these high values of $\chi_{\nu}$ is scattered widely in radius.
We therefore make allowance for this scaling the statistical
errors up by the factor required to give
$\chi_{\nu}=1$, and then use these larger errors to perform the error
calculation (using $\Delta\chi^{2}=1$) for the fitted parameters. 
This allows for the effects of the excess scatter in weakening
the constraints on our derived parameters. The
errors derived in this way for the global fit are included in Table~3.

Inspection of Fig.~2 suggests that spherically symmetric
$\beta$-models might not be good representations of the surface
brightness distributions of the two clusters. For example, it can be
noted in the same figure that the cD in Abell~399 is not at the centre
of the X-ray surface brightness distribution, which additionally does
not appear spherically symmetric, especially at large radii. On the
other hand, Abell~401 is elongated along the NorthEast-SouthWest
direction. In order to disclose these anomalies, we perform each
image analysis in four sectors around the central cluster
galaxy.

\subsubsection{Sectors}

In the previous section we have found that the surface brightness
distribution of both clusters can be described by flat
$\beta$-models. In addition to this component, some central excess is
required.  Given the apparent proximity of the two clusters to each
other, we might expect that tidal interactions have extended the gas
distributions in each towards its neighbour. This effect would
manifest itself in azimuthal differences in the surface brightness
distributions at large radii. To explore this expectation, we model
the radial cluster profiles in four 90~deg sectors for each cluster,
centered, as in Section~3.1.1, on the peak of the X-ray emission which
coincides with the cD galaxy. The sectors we use are shown in
Fig.~\ref{A399pies}, and are labelled N, W, S and E for each cluster.
We accumulate counts in annular segments of width 30~arcsec, and each
radial profile is fitted by a single $\beta$-model, as in Section~3.1.1, and
the results listed in Table~3.

For clarity, in Fig.~\ref{A399T_prof_pies}, and \ref{A401T_prof_pies}
we compare the surface brightness profiles in the four sectors for
each cluster with the $\beta$-models found from the global fits to the
data with a single model. These figures also plot the temperature
profiles found in the same sectors and their derivation and discussion
will be presented later in this paper (Section~3.2.3).

\begin{figure*}
\begin{center}
\leavevmode 
\epsfxsize 0.8\hsize
\epsffile{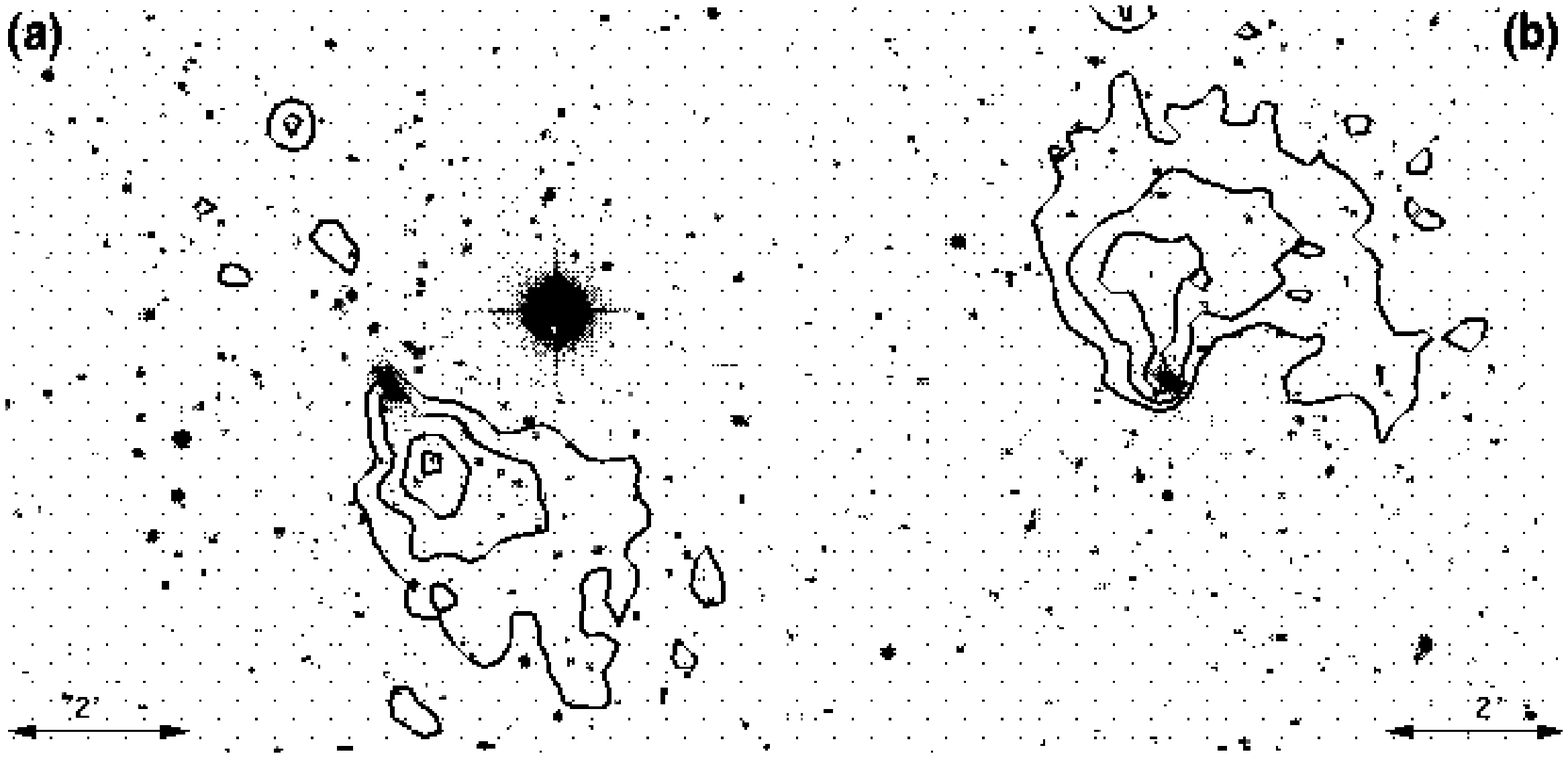}
\end{center}
\caption{Positive residuals of the 2-dimensional fit overlaid onto a DSS image: (a) Abell~401, and (b) Abell~399. The lower contour lever is
approximately at 3$\sigma$ of the surrounding `background' level. The
image scale is also shown.}\label{RES}
\end{figure*}

Inspection of Fig.~\ref{A399T_prof_pies}, and \ref{A401T_prof_pies}
reveals that the clusters are not azimuthally symmetric.  One of the
most intriguing features that arise from the above comparison is that
at large distances from the cluster centres, the ICMs appear more
extended only in the sectors that face the other member of the binary
system. Additionally, Abell~399 is more extended also to the West. For
Abell~399 we find the $\beta$ index to lie between 0.497 (sector-S)
and 0.512 (sector-N), straddling the values found from the azimuthally
average analysis. The range of $r_{\rm c}$ is wider. Sector-E gives
the smallest ($r_{\rm c}$=0.997~arcmin), and sector-W the largest
($r_{\rm c}$=3.129~arcmin). The values for sector-N, which is the one
that is facing Abell~401, are average between the quoted extremes.
Excluding from the fit the inner points does not make any difference
to the best fit values.

Although Abell~401, looks more symmetric than Abell~399, the range of
$\beta$ values we find is larger. Sectors-S and E give the smallest
$\beta$ (0.545 and 0.515 respectively). The fit to the data from the
sector that is the furthest away from Abell~399 (sector-N) results in
the steepest $\beta$ of 0.727, and the largest $r_{\rm c}$ of
2.716~arcmin. The core radius of the sector that is facing Abell~399
(sector-S) is average ($r_{\rm}$=1.937~arcmin).

To summarise the results of sections Section~3.1.1 and Section~3.1.2, we find
$\beta$ values in the range 0.50-0.57 in Abell~399, and 0.50-0.73 in
Abell~401. These values are smaller than the `canonical' value of 0.65
for clusters of galaxies.  Such small $\beta$ values have been found
before in binary and merging clusters of galaxies (e.g., Donnelly et
al. 2001). Only the data from sector-N and -W in Abell~401, which are
on the opposite side of the other cluster, have larger $\beta$s of
0.73 and 0.65 respectively. However, as we will demonstrate later
(Section~4), the North sector in Abell~401 is unusual also in the
temperature profile.  Abell~399 appears more disrupted, and as can
also been derived from Fig.~\ref{A399T_prof_pies}, and Table~3, the
surface brightness distribution in almost all sectors is extended,
with small $\beta$s. Sector-E is abnormal giving a small core radius
of $r_{\rm c}$=0.997~arcmin=81.35~kpc. Additionally, we find that both
clusters show a central excess which is more apparent in Abell~399,
and we can confidently exclude the possibility that this is due to any
active nuclei that reside within the core of the clusters.

\subsubsection{2-dimensional residuals}

In order to investigate the origin of the central excess in both
clusters seen in the surface brightness profiles, and visualise the
azimuthal differences that the results of Section~3.1.2 suggest, we perform
2-dimensional fits to the image data for both clusters. We use the
background subtracted and exposure corrected images discussed above,
for this spatial analysis, and fit the images of each instrument and
cluster individually in {\sc sherpa}, using 2-dimensional
$\beta$-models. The values of $r_{\rm c}$ and $\beta$ are fixed to the
best fit values found in the azimuthal average 1-dimensional fits with
a single $\beta$-model. The centres of the 2-dimensional models are
fixed to the values used for the 1-dimensional analysis, which
coincide with the location of the central galaxies. The normalisations
of the three instruments are left free to vary.  The residual images
for the MOS cameras are co-added and smoothed using the SAS task {\sc
asmooth}. Figure~\ref{RES} shows contour plots of the residual
smoothed images overlaid onto DSS images of the central regions of
both clusters.

A few new intriguing features appear from the above 2-dimensional
analysis, which are not very apparent in the 1-dimensional
plots. Firstly, in both clusters, the positive residuals lie on one
side of the central galaxy: to the South of the galaxy in Abell~401,
and to the North of the one in Abell~399 with some additional
extension to the west. We expect, though, that some of the residuals
in Abell~401 stem from the fact that we fit a circular symmetric model
to a cluster that is intrinsically elongated along the same direction
as the excess seen in Fig.~\ref{RES}. However, if this is the only
reason for the appearance of the residuals in Abell~401, it cannot
explain why we don't see similar residuals on the opposite side of the
galaxy to the North-East. If we leave the ellipticity and position
angle of the 2-dimensional $\beta$-model free, we obtain an
ellipticity of $e$=0.115, a position angle of 2.0. The residuals are
very similar, being on the South-West side of the central galaxy
only. Similarly, the 2-dimensional fit to the Abell~399 images reveals
that the central excess seen in the 1-dimensional profiles of
Fig.~\ref{profiles} can be attributed to emission coming from one side
of the galaxy, mainly the North and West.

Thus, the 2-dimensional analysis reveals the shape of the central
excesses in both clusters, which were noted in the 1-dimensional image
analysis. The positive residuals of Fig.~\ref{profiles} are prominent
in the innermost regions of the clusters, extending
(2-3)~arcmin$\simeq$ (170-250)~kpc around the central galaxy. As seen
in Fig.~\ref{A399T_prof_pies} and \ref{A401T_prof_pies} these
extensions appear to carry on on larger scales.

\subsection{Spectral Properties}

For any following spectral analysis, we use the clean event lists for
the data and background. We impose further restrictions by keeping
events that have {\sc flag}=0 and {\sc pattern}=0. This filtering
ensures us that the events we use for the spectral analysis have
cosmic origin, and they are not generated by for example cosmic rays,
or CCD defects. Responses and auxiliary files are generated with {\sc
rmfgen-1.44.6 } and {\sc arfgen-1.48.10} respectively.

For the fitting procedure, we only use the (0.6-7.5)~keV range for the
observation 301 (Abell~401). The reason we increase the low energy
limit is because it was taken with the medium filter, while
for all the blank-sky background files the thin filter was used. The
effective areas of the thin and medium filters mostly differ at low
energies ($<$1~keV; \xmm User-Handbook). Thus, in order to avoid inaccurate
subtraction of the background at low energies, we exclude energies
below 0.6~keV. This cut-off may cause some overestimation of
temperatures by a small amount, and/or inaccurate determination of the
absorbing column density. However, given that Abell~401 is a hot
cluster, we don't expect to introduce any significant bias
towards high temperatures.  Several tests with and without the
(0.3-0.6)~keV energy range show that the temperatures calculated with
and without the increased low energy limit agree within the
errors. For the sake of consistency between the different
observations, we decide to use the same energy range (0.6-7.5)~keV for 
any subsequent spectral analysis. Another reason that this selection
is desirable, is that calibration problems that are mostly prominent
at low energies are avoided.

In the following sections we present the results of the spectral
analysis for both clusters.

\begin{table}
\begin{center}
\caption{Results: Spectral Fits}\label{spectra_info}
 \begin{tabular}{ccccccccc}   \hline \hline

Cluster		&
$N_{\rm H}$		&
$kT$		&
$Z$		&
$\chi^{2}/d.o.f.$
\\

		&
($\times 10^{21} \ {\rm cm^{-2}}$)		&
(keV)		&
(${\rm Z_{\sun}}$)		&

\\

\hline

A399		&
0.812$_{-0.017}^{+0.047}$		&
7.23$_{-0.21}^{+0.17}$ 		&
0.22	$_{-0.02}^{+0.02}$	&
1335/930
\\

A401		&
0.729$_{-0.005}^{+0.041}$		&
8.47$_{-0.38}^{+0.12}$		&
0.25$_{-0.02}^{+0.02}$		&
1445/1203
\\

\hline

\end{tabular}
\vspace{0.2cm}
\begin{minipage}{8cm}
\small The errors quoted are the 90~percent errors for one parameter of
interest. Solar abundances were taken from Anders \& Grevesse (1989).

\end{minipage}
\end{center}
\end{table}

\begin{figure}
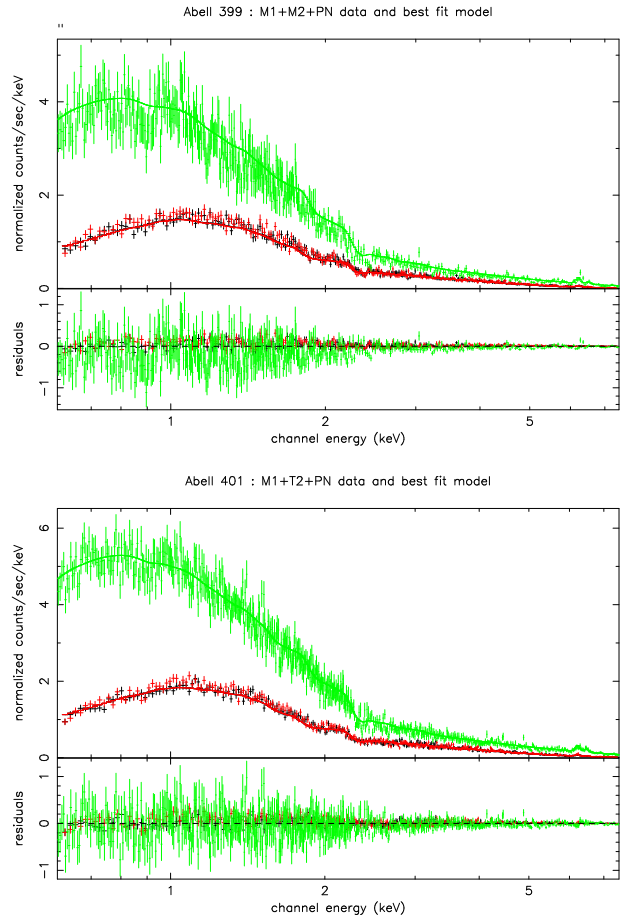
\label{spectrum}
\begin{center}
\setlength{\unitlength}{1cm}
\begin{picture}(8,7)
\put(-0.5,8){\includegraphics{A399_all_spec.ps}}
\end{picture}
\end{center}
\begin{center}
\setlength{\unitlength}{1cm}
\begin{picture}(8,5)
\put(-0.5,7){\includegraphics{A401_all_spec.ps}}
\end{picture}
\end{center}
\caption{\xmm spectra of the central regions of Abell~399 (top panel)
and Abell~401 (bottom panel). MOS~1, MOS~2, and PN spectra are
accumulated from the inner 10~arcmin region around each cluster
centre.  The best fit model and residuals of the fit are also
shown. data from the two MOS cameras are shown as black and grey
crosses, and give the same count rate. The spectra accumulated from
the PN instrument are always the ones at higher ${\rm cnt \ s^{-1} \
keV^{-1}}$.} 
\end{figure}

\subsubsection{Global Spectral Properties}

Spectra of both clusters are accumulated over a circular region with
radius of 10~arcmin around each cluster centre. Background spectra
are taken from the blank-sky background files in the same regions on
the detectors as for the data spectra. The background spectra are scaled by the factors found in Section~2.2. We generate spectra from each
instrument independently, and subsequently fit all three instruments
in {\sc xspec} simultaneously.

The spectra are fit with a {\sc mekal} model modified by the
line-of-sight hydrogen absorption, as described by the {\sc xspec}
{\sc wabs} model. The galactic absorption for the direction of both
clusters is $N_{\rm H,G}=1.03 \times 10^{21} \ {\rm cm^{-2}}$. During
each fitting procedure, the hydrogen column density ($N_{\rm H}$),
metal abundance ($Z$), temperatures ($kT$) and normalisations are left
as free parameters.  Their best fit values are tabulated in
Table~4, along with the goodness of each fit.  In
Fig.~7(a) and Fig.~7(b) we present the
spectra of Abell~399 and Abell~401 respectively. The best fit thermal
models, and the residuals of the fits are also shown in the same
plots.

For both clusters, we find an absorbing column consistent with the
galactic value. The global temperatures and abundances are in good
agreement with previous results from the {\it ASCA} satellite
(Nevalainen et al. 1999; Markevitch et al. 1998; Fujita et al. 1996).

\subsubsection{Temperature Profiles}

\begin{figure*}
\begin{center} 
\leavevmode 
\epsfxsize 0.45\hsize
\epsffile{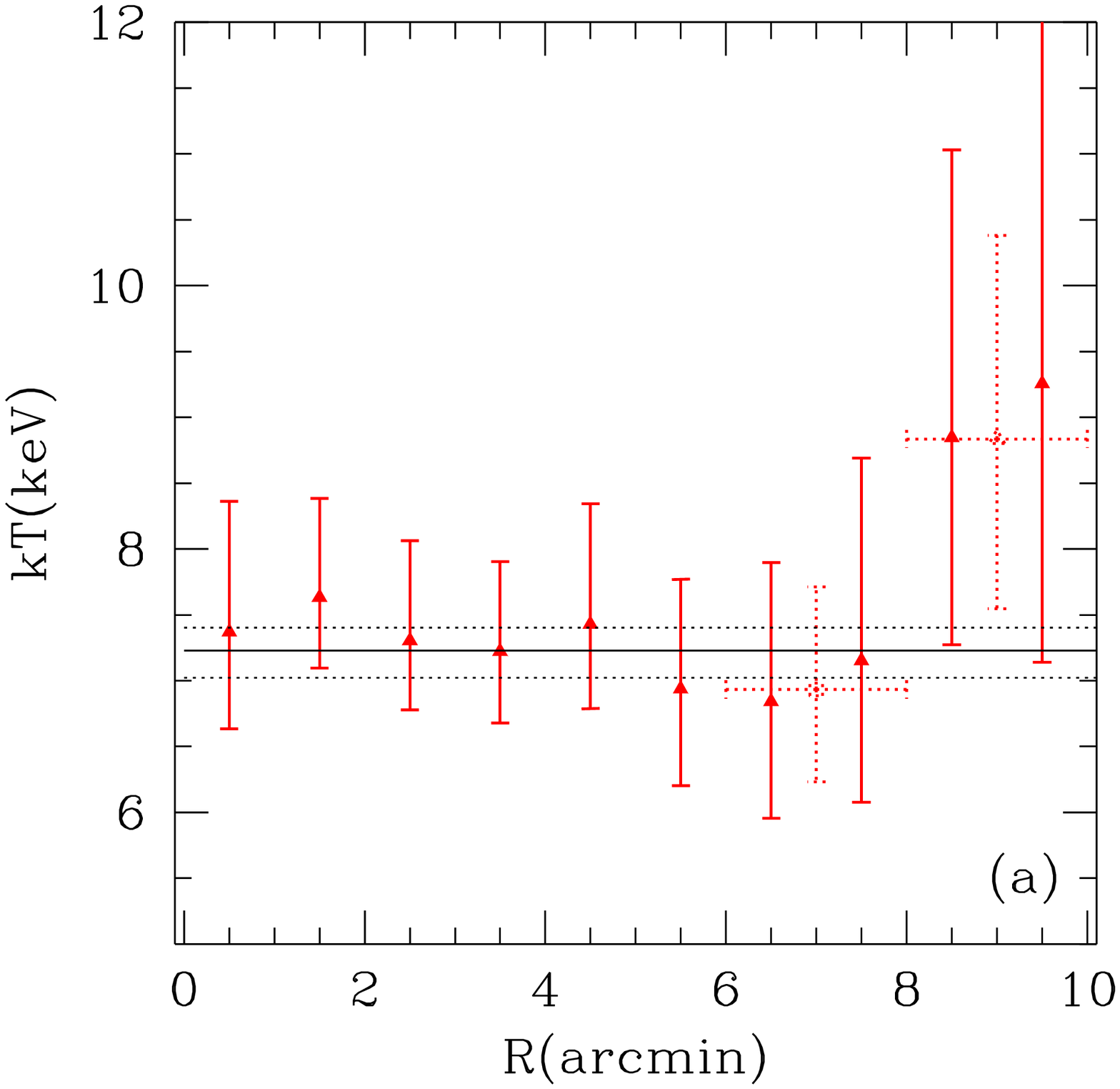}
\leavevmode 
\epsfxsize 0.45\hsize
\epsffile{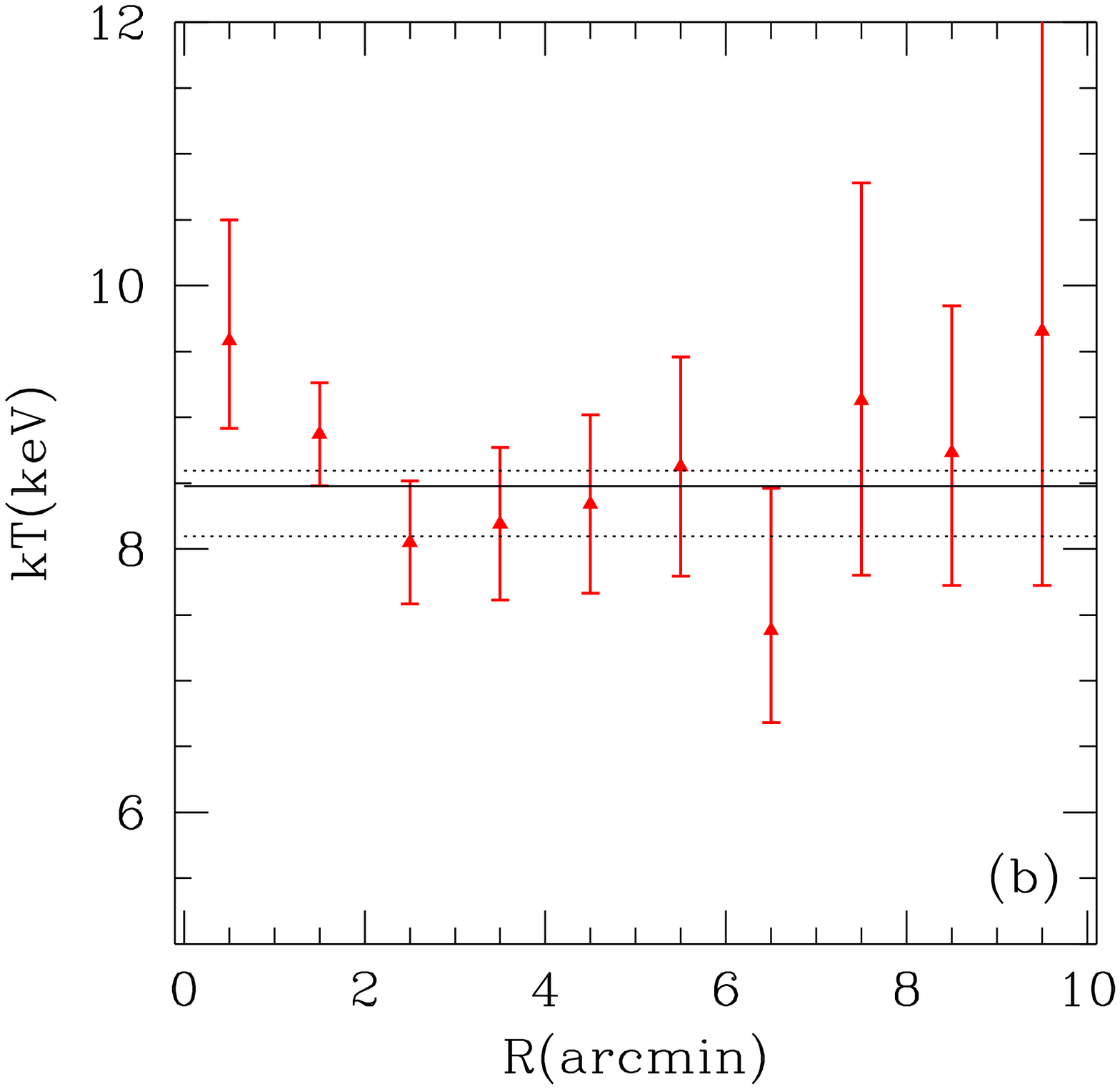}
\end{center}
\caption{Projected temperature Profiles for Abell~399 (left panel) and
Abell~401 (right panel). The width of each annulus is 1~arcmin (solid
lines). The derived temperature in 2~arcmin wide bins for the regions
from 6 to 10~arcmin is shown with dotted lines.  
The global temperatures (solid horizontal line) and their confidence
limits (dotted horizontal lines) are also shown.}\label{T_prof}
\end{figure*}

Assuming spherical symmetry, we obtain radial temperature profiles, by
accumulating source counts in circular annuli around each cluster
centre.  The width of each annulus is 1~arcmin. As for the global
spectral analysis, we fit the (0.6-7.5)~keV energy range.  The
background is taken again from the blank-sky clean and filtered files,
as in Section~~3.2. During the fitting procedure, the absorbing column
$N_{\rm H}$ is fixed to the best fit value for each cluster listed in
Table~4. The abundances are left free to vary.  In the inner 7 annuli,
after background subtraction there are more than 80~per cent of the
total counts left in the energy range we use for the fits. Only the
last 1-2 annuli, are left with (65-75)~percent counts after the
background subtraction.  The reduced $\chi^2$ varies between
$\chi^{2}_{\nu}$=0.55-1.15 for Abell~399, taking these extreme values
at 0.5 and 9.5~arcmin respectively, while the rest give values of
$\chi^{2}_{\nu}$ between 0.94 and 1.13. For Abell~401, the fits are
generally better, with $\chi^{2}_{\nu}$ lying between 0.93 and 1.12
(for the annuli at 4.5 and 1.5~arcmin
respectively). Figure~\ref{T_prof} shows the resultant projected
temperature profiles, for Abell~399 [Fig.~\ref{T_prof}(a)], and for
Abell~401 [Fig.~\ref{T_prof}(b)]. In both clusters the calculated
abundances are consistent with their global values.  Both clusters
appear nearly isothermal in their inner regions, and the derived
temperatures are consistent with the average values of Table~4.

We have also derived deprojected temperature profiles. As expected,
these are similar to the projected ones, since there is no strong
gradient in the temperature. Of course, the deprojection procedure has
the effect of increasing the errors of the temperatures.

As can be seen in Fig.~\ref{T_prof}(a), Abell~399 appears isothermal
at the global temperature out to $\sim$8~arcmin = 0.65~Mpc. At large
radii [(8-10)~arcmin], the spectral fits give higher values for the
temperature. Past results for the temperature slope of clusters of
galaxies at large radii derived from {\it ASCA} data (e.g., Markevitch
et al. 1998) and {\it ROSAT} data (Irwin et al. 1999) led to
contradictory results for the temperature gradient at large radii. The
general trend is that {\it ASCA} finds a decline in the temperature
profiles at large radii, while with {\it ROSAT} clusters appear at a
nearly constant temperature. The past {\it ASCA} results for Abell~399
are conflicting too, but this disagreement could be understood in
terms of the different treatments for {\it ASCA}'s PSF, by different
authors. For example, Markevitch et al. (1998) found that the
temperature remains constant at $\sim$6.2~keV between 3 and 16~arcmin,
slightly lower than our values, but consistent within the errors. They
also found a central excess, with the temperature within the inner
3~arcmin, escalating up to 8.5~keV, which is not evident in the \xmm
results of Fig.~\ref{T_prof}(a). On the other hand, Fujita et
al. (1996) did not see any central excess, and argued that the cluster
is nearly isothermal out to 10~arcmin. At large radii, they found some
evidence for a small temperature increase in their data. A temperature
increase at $\sim$10~arcmin was also evident in the temperature
profile presented in White (2000).
 
Thus, up to now, there is no consistent result for the temperature of
the cluster's ICM at $\sim$10~arcmin from the centre of Abell~399. As
noted before, we find a small increase at such large radii [see
Fig.~\ref{T_prof}(a)]. In order to investigate whether this apparent
increase is due to the low number of counts from the cluster at these
large radii, we increase the size of the last annular bin, and we
extract the spectra in annuli between 6-8 and 8-10~arcmins. The
derived temperatures are shown as dotted crosses in
Fig.~\ref{T_prof}(a). The calculated $\chi^{2}_{\nu}$ for both are
1.07 and 0.90. As can be seen the results are consistent with the
results from the 1~arcmin wide annuli, showing a small temperature
increase.

Another possibility for the derived hotter temperatures, could be that
the scaling of the blank-sky background files was not adequate, but
the background level needs to be increased further. To test this
possibility, we increase the background by a factor of 1.2 and perform
the same spectral fits to the wide annuli as before. The derived
temperatures are $6.45^{+0.73}_{-0.67}$~keV and
$7.61^{+1.46}_{-0.95}$~keV for the inner [(6-8)~arcmin] and the outer
[(8-10)~arcmin] annuli respectively. Thus, a 20~percent increase of
the background level results in the reduction of the temperature by
$\sim$14~percent, from a temperature of
$8.84^{+1.55}_{-1.29}$~keV. This scaling reduces the temperatures of
the inner annuli by a small amount, with a decrease of $\sim$4~percent
the maximum.  Increasing the background scaling factor more leads to
worse fits. Generally, the models do not represent the data at
energies $<$1~keV. More absorption is required at these low energies
to compensate the increased flux from the lower temperature plasma, if
the temperature is to be lower than the one shown in
Fig.~\ref{T_prof}(a). There is no obvious reason to support an
increase of the absorbing column density with radius from the cluster
centre.  Thus, we conclude that an increase of the scaling factor of
the background level by a factor not larger than 1.2 could suppress
the high temperatures found by the previous analysis at large radii,
without changing much the temperature of the inner annuli. However, we
are not aware of such large scaling factors required in any other data
sets, and we decide to use the factors found in Section~2.2, and state
where necessary the effect of an increased value to the derived
temperatures.

On the other hand, Abell~401 shows isothermality at the global
temperature out to $\sim$10~arcmin=0.83~Mpc [see
Fig.~\ref{T_prof}(b)]. There is also some evidence for a central
temperature increase in the inner $\sim$2~arcmin, as previously noted
in the {\it ASCA} data (e.g., Nevalainen et al. 1999).  Previous
analysis of the {\it ASCA} data indicate that the ICM at a distance of
(8-16)~arcmin might be cooler than in the inner regions by a factor of
$\sim$0.6 (Nevalainen et al. 1999).

\subsubsection{Sectors}

In order to disclose any azimuthal dependences of the temperature
profiles, we obtain the spectra in concentric annuli in the four
sectors shown in Fig.~\ref{A399pies}, as for the spatial analysis. We
now increase the width of each annulus, so that the number of counts
recorded in each one is adequate for the full spectral modelling. We
perform exactly the same fitting procedures as in Section~3.2.2. The derived
temperature profiles are shown in Fig.~\ref{A399T_prof_pies} for
Abell~399 and Fig.~\ref{A401T_prof_pies} for Abell~401. The same
figures show the surface brightness distributions in the sectors, and
were discussed earlier in this paper.  

A few striking properties for the temperature structure in Abell~399
emerge from these figures: i) the high temperature in the inner
2~arcmin in sector-S, ii) the high temperature in the outer
(6-10)~arcmin region in sector-E, iii) along sector-N, towards
Abell~401, the temperature profile appears the most regular of the
four, showing the gas to be isothermal at the global temperature.

\begin{figure*}
\begin{center}
\leavevmode 
\epsfxsize 0.45\hsize
\epsffile{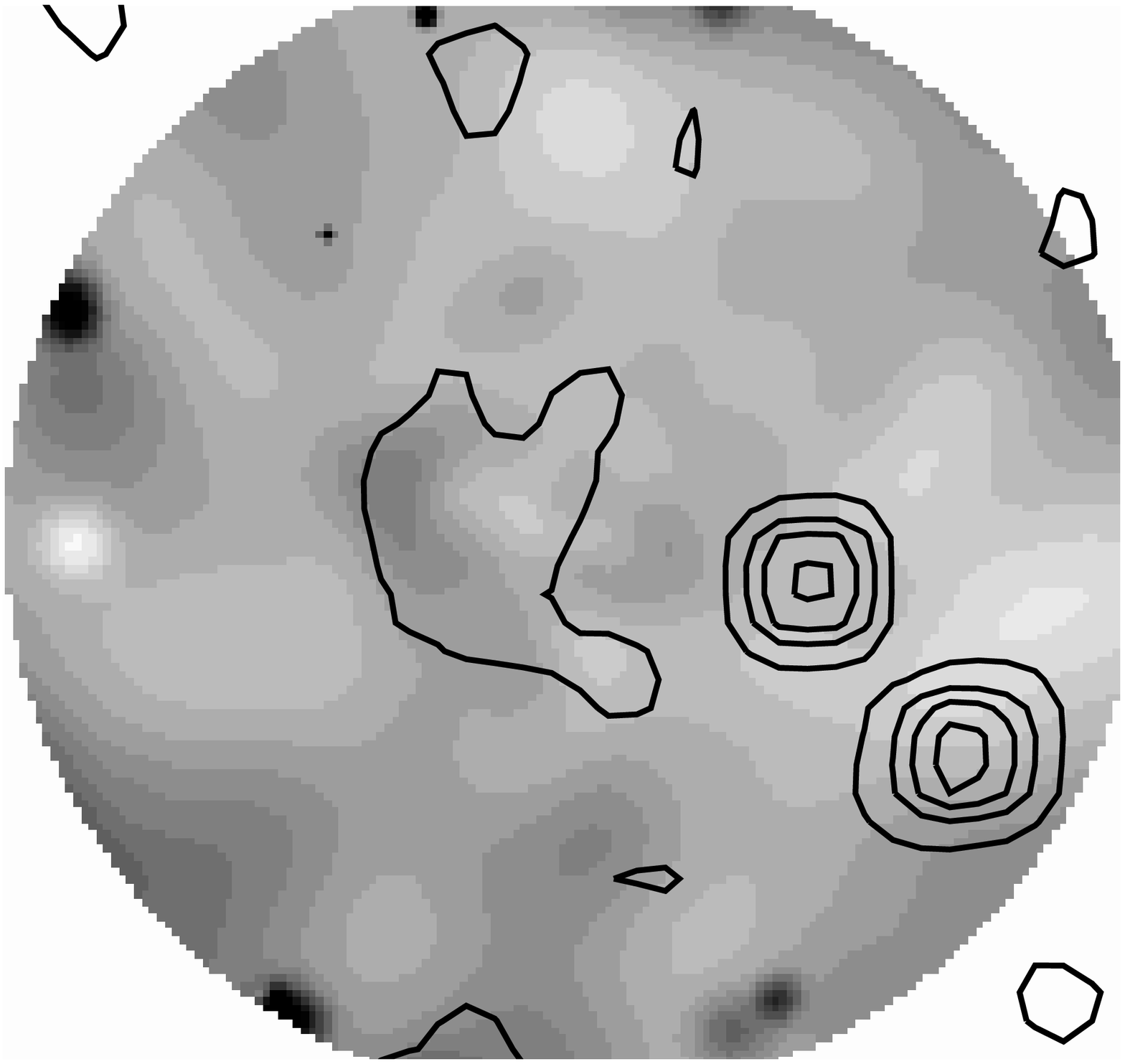}
\leavevmode 
\epsfxsize 0.45\hsize
\epsffile{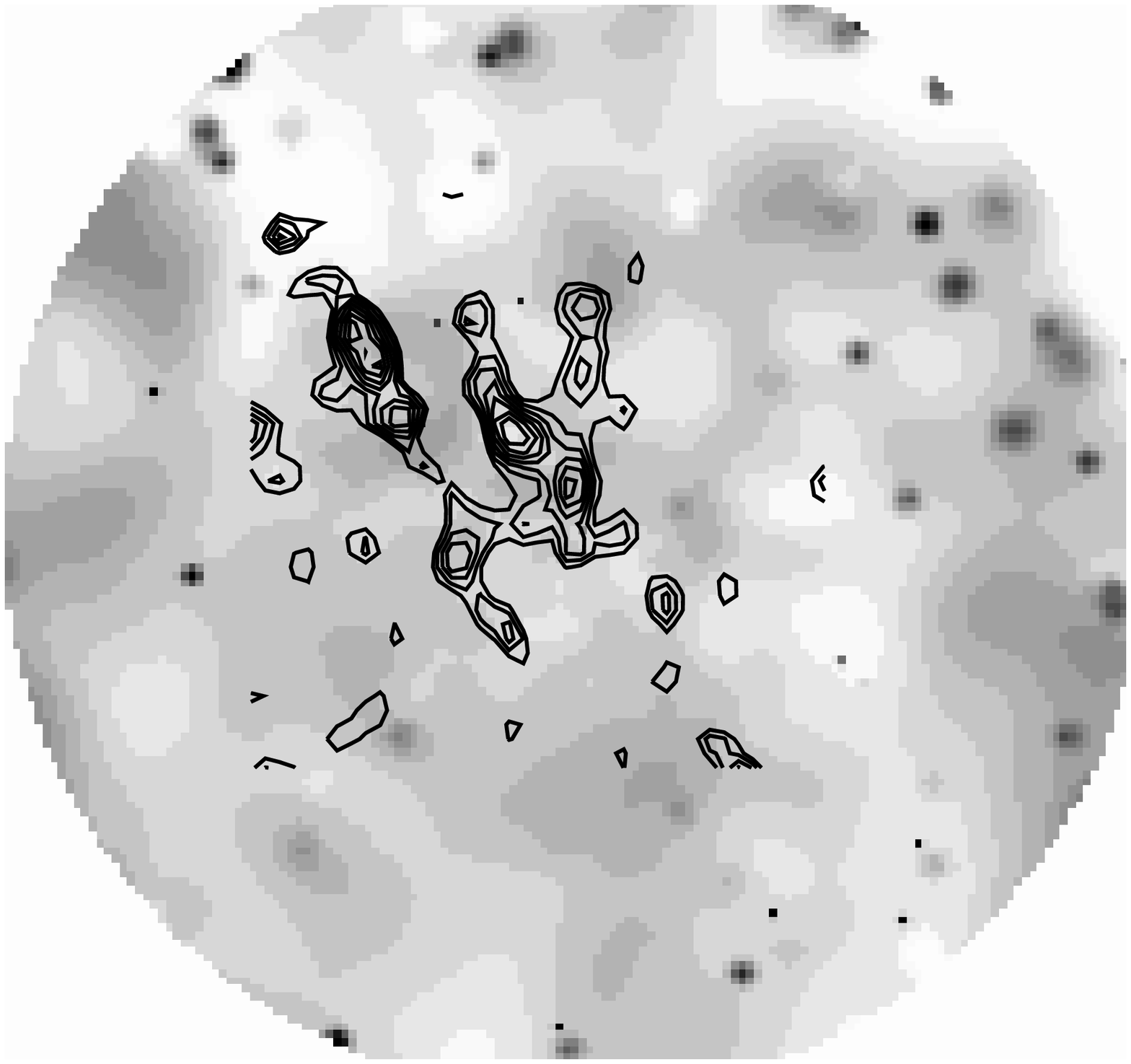}
\end{center}
\caption{Hardness ratio maps of Abell~399 (left panel) and Abell~401
(right panel). On both maps we overlay the radio contours, taken from
the NVSS (see Section~6). The radii of the circular regions are
10~arcmin in size,
and their centres coincide with the location of the cD galaxies and
the cluster centres. Harder emission is shown with darker shades of
grey, and the scale is arbitrary.}\label{HR}
\end{figure*}

The high temperature found in the outer annuli in sector-E should be
the main reason for the increased temperature found in the azimuthally
symmetric temperature profiles of Section~3.2.2. As seen in the raw \xmm
images, in this sector at large radii the flux from the cluster is
rather low, making the spectra more depended on the background
level. If we increase the scaling of the background by a factor of
1.2, as in Section~3.2.2., we find a temperature for the last annuli in
sector-E of 8.81$_{-2.30}^{+4.99}$~keV, compared to a temperature of
11.48$_{-2.80}^{+4.94}$~keV using the original scaling. Thus, an
increase of the scaling factor by the additional factor of 1.2,
reduces the temperature by almost 25~percent. A better estimation of
the temperature along sector-E at large radii will have to wait for
data with longer exposures that compensate for the inherently low flux
from this sector.

Abell~401 also shows some small scale irregularities in the
temperature profiles of Fig.~\ref{A401T_prof_pies}, i) the sector that
is facing Abell~399 appears the most regular, nearly at the global
temperatures, ii) the inner 2~arcmin of sector-N and -W are slightly
hotter than the rest, and are probably responsible for the central
temperature increase noted in Fig.~\ref{T_prof}(b), iii) there is a
temperature drop out to large radii along sector-N. This property
will be discussed later in this paper (Section~5).  
 
\subsection{Central Regions}

Past X-ray data showed no significant evidence for the presence of
cool gas in either cluster core (e.g., Peres et al. 1998). Our higher
quality data confirm the absence of centrally concentrated cold
material.  We have found that a single temperature model gives an
adequate representation of the spectra in the inner 1~arcmin regions,
leaving no room for any extra thermal or non-thermal components.
Trials with cooling flow models fail to provide acceptable fits. Thus,
we conclude that none of the two clusters hosts even a mild cooling
flow in its centre. Additionally, this result supports the statements
of Section~3.1.2., that the central excess seen in the surface brightness
profiles are not due to an active galaxy, since there is no spectral
evidence for it.

\begin{table}\label{summary}
\begin{center}
\caption{Results: Key Properties}\label{spectra_info}
 \begin{tabular}{lll}   \hline \hline

Property		&
Abell~399		&
Abel~401		
\\
\hline

$kT$			&
7.23			  &
8.47
\\

(keV)			&
			&
\\

$r_{\rm c}$		&
155.3			&
169.6
\\

(kpc)			&
			&
\\

$\beta$			&
0.50			&
0.59
\\

$n_0$			&
4.19			&
6.76
\\

($\times 10^{-3} \ {\rm cm^{-3}}$)	&
					&
\\

$t_{\rm cool}$				&
18.8					&
12.4
\\

(Gyr)					&
					&
\\

$M_{500}$		&
4.98			&
6.13			
\\
$(\times 10^{14} \ M_{\sun}$)	&
					&
\\

  $R_{200}$		&
2.16			&
2.34
\\
(Mpc)			&
			&
\\

\hline

\end{tabular}
\end{center}
\end{table}

\begin{figure*}
\begin{center} 
\leavevmode 
\epsfxsize 0.45\hsize
\epsffile{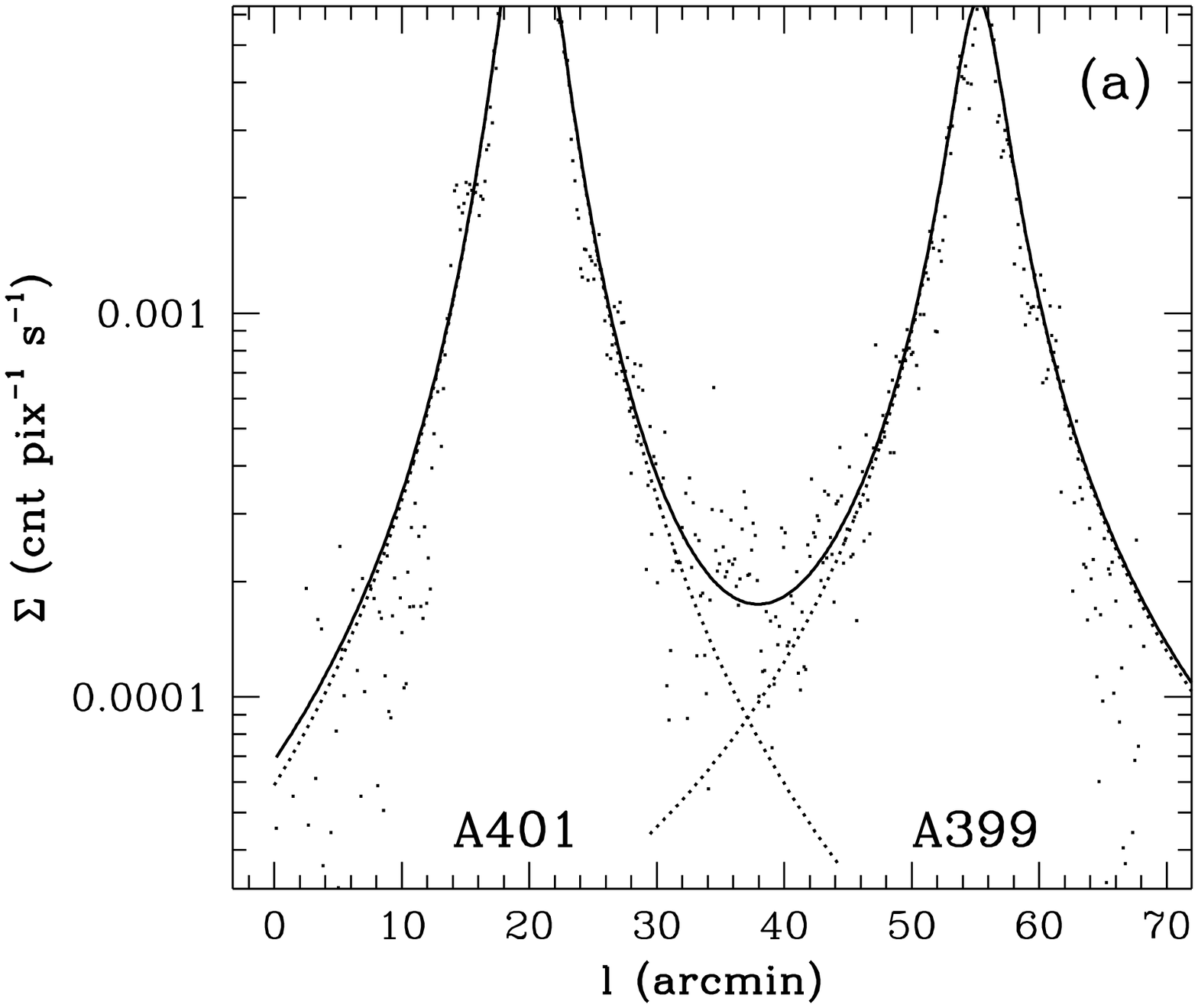}
\epsfxsize 0.45\hsize
\epsffile{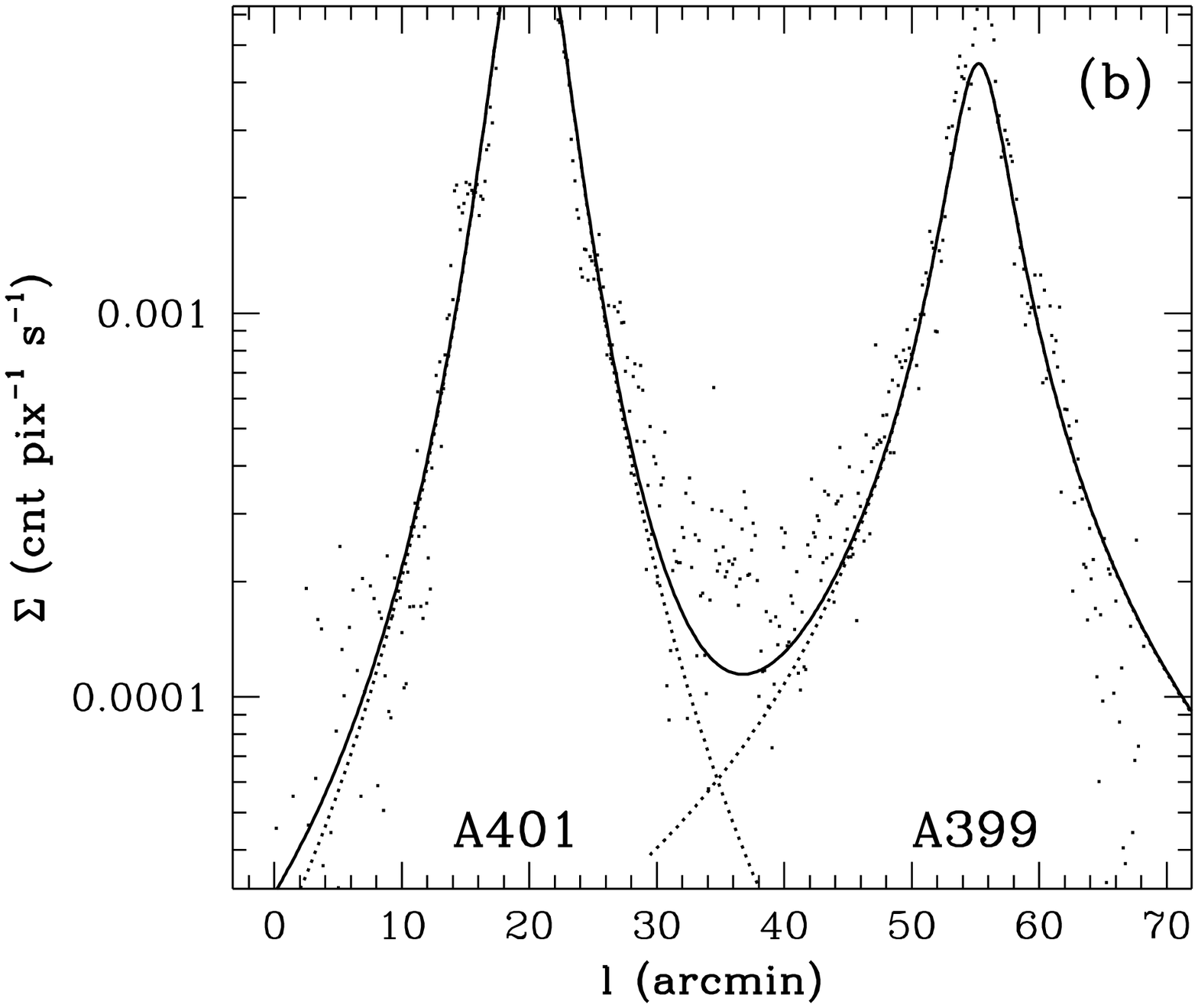}
\end{center}
\caption{Surface brightness profiles along a 1~arcmin wide strip that
intersects the two cluster centres. The best fit $\beta$-models
derived in Section~3.1 are shown with dotted lines. In the left panel we
present the global $\beta$-models, and in the right panel the best fit
models of the sector-N for Abell~401, and sector-S for Abell~399 are
used (see text for more details). The superposition of the two cluster
models is shown with the solid line.}\label{cuts}
\end{figure*}

\subsection{HR maps}

In Fig.~\ref{HR} we present the hardness ratio images of the central
10~arcmin for both clusters. These maps are obtained by dividing a
hard image ($H_{im}$) by the soft ($S_{im}$). The energy ranges for
the soft and hard are $S_{im}$=(0.5-1.0)~keV, and
$H_{im}$=(2.0-5.0)~keV. We avoid the (1.0-2.0)~keV energy range as it
contains strong background lines. The $S_{im}$, and $H_{im}$ are
corrected for exposure, and a background is subtracted, in a similar
manner as in Section~3.1. Before we divide the images, we adaptively smooth
them, using the smoothing scales derived for the hard image. On the
hardness ratio maps we overlay the radio contours, which show the
location and extent of the clusters' radio halos (see Section~6 for more
details).

These images show the small scale variations noted in the temperature
profiles of Section~3.2.3. For example, the region in the (2-4)~arcmin in
the east sector in Abell~399 seems to coincide spatially with the
extended radio emission from this cluster. The emission from the north
of Abell~401 appears softer than elsewhere in the cluster, confirming
the results of the spectral fits of Section~3.2.3.

\subsection{Summary of Global Results}

In Table~5. we collect some key  properties of both clusters,
derived from the \xmm analysis. The central number density of the ICM
is calculated by deprojecting the central surface brightness
($\Sigma_{0}$) found from the spatial analysis of Section~3.1, using eq. (3)
and (4) from Sakelliou et al. (1996).  The cooling time $t_{\rm
cool}$, which can be treated as an indication of the presence of a
cooling flow (Sarazin 1986), is also given in Table~5.  The number
density $n_0$ and cooling time $t_{\rm cool}$ given are their central
values.  For the $M_{500}$ we follow Finoguenov et al. (2001), and for
$R_{200}$ the results of Evrard, Metzer \& Navarro (1996) are used.

\section{Between Abell~399 and Abell~401}

We investigate next the properties of the region between the two
clusters. The separation of the cores of the two clusters is
$\sim$3~Mpc, and their $R_{\rm 200}$, as calculated from their global
temperatures found in Section~3.1.1 are presented in Table~5. In this binary
system, the $R_{\rm 200}$ surfaces of both clusters overlap (unless
there is a large projected separation between them, which is not
justified by the redshift data, as mentioned in the introduction),
making it difficult to believe that the properties of each is not
influenced by the presence of the other. If they have started merging
with each other, or have been through each other already, one would
expect the characteristics of the gas between them to show the
signatures of these physical processes, and carry some clues about the
dynamical state of the system.

As was mentioned above, one of our \xmm pointings (the 201
observation) was centred between the two clusters, and we use it to
obtain the X-ray properties of the gas occupying this
region. Background subtracted and exposure corrected images are
created in a similar way as in Section~3.1, and for the spectral analysis we
follow the analysis of Section~3.2.

\subsection{Surface brightness}

In Fig.~\ref{cuts} we show the count flux along a strip of 1~arcmin
width, that passes though the two cluster centres. In
Fig.~\ref{cuts}(a) we plot the best fit models derived from the fits
to the azimuthal average surface brightness profiles of Section~3.1.1.  In
the same plot, we also show the superposition the two $\beta$-models
with the solid line. Two important conclusions can be drawn from
inspection of this plot.  Firstly, the azimuthal average profiles do
not represent well the data along the narrow strip, but lie
systematically above the data at large radii. These negative residuals
in respect to the global fits are apparent Fig.~\ref{A399T_prof_pies}
(sector-S), and in Fig.~\ref{A401T_prof_pies} (sector-N).  Thus, the
North-South direction of the narrow strip should exaggerate these
discrepancies. In order to try and take into account this behaviour,
we plot in Fig.~\ref{cuts}(b) the $\beta$-models derived from the
North sector for Abell~401 (sector-N) and the South one for Abell~399
(sector-S), assuming, of course, mirror symmetry about the cluster
centre. It is now seen in Fig.~\ref{cuts}(b) that these models
represent better the surface brightness distribution at large radii to
the north of Abell~401. However, the model from sector-S in Abell~399
is still not a good representation of the profile at radii
$>$10~arcmin to the South of Abell~399. Unfortunately, we do not have
another pointing to the South of Abell~399, which could check
accurately the distribution of the cluster at such radii. Given that
the discrepancy is mainly at radii $>$10~arcmin, where the
uncertainties of the vignetting function and the background
subtraction become more important, we use the model presented in
Fig.~\ref{cuts}(b) for the following discussion.

In Fig.~\ref{cuts}(b) there is evidence for an excess between the two
clusters, which cannot be explained solely by a simple superposition
of the flux from the two clusters. In this region we find an average
flux above the background of 0.25~$ \times 10^{-3}~{\rm cnt \ s^{-1} \
pix^{-1}}$, while the combined model predicts a flux of
$\sim$0.13~$\times 10^{-3}~{\rm cnt \ s^{-1} \ pix^{-1}}$.

\subsection{Temperature}

\begin{figure}
\begin{center} 
\leavevmode 
\epsfxsize 0.95\hsize
\epsffile{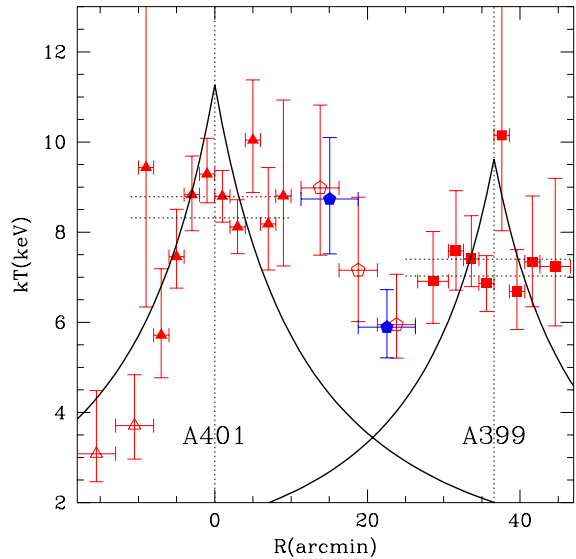}
\end{center}
\caption{Temperature variations along a line that bisects the two
cluster centres. Abell~401 is at R=0 and Abell~399 is at R=36.6. The
temperatures for Abell~401 are shown with triangles, and for Abell~399
with squares. Open triangles mark the temperatures found from the 401
observation, while solid triangles come the 301 one (sectors-N and -S
only). Similarly, the temperature variations in Abell~399 (sector-N
and -S) are shown. The temperatures for the space between the two are
shown with pentagons (open for the three 5~arcmin wide boxes, and
filled for the two 7.5~arcmin boxes). The global cluster temperatures
found in Section~3.2.1 are marked with the dotted horizontal lines, and the
perpendicular ones show the locations of the cluster centres. Solid
lines plot the temperature profiles predicted for Abell~401 and
Abell~399 using the universal profile of Loken et al.
(2002).}\label{T_A399401}
\end{figure}

There are various plausible explanations for the presence of extra gas
in-between the clusters. It could be just the ICMs of the two
clusters, that have been stretched due to the action of gravitational
forces. Alternatively, it could be a compression region, that is
created as the two clusters come together. Thirdly, it could be
cluster gas left behind after the two cluster cores have passed each
other. All these processes should leave different signatures in the
spectral properties of the gas. If for example, we are seeing a
compression region, its temperature should be higher than the
surroundings. In order to investigate this issue we examine the
variations in temperature along the line that connects the two cluster
centres.  Due to the low number of counts in the region between
Abell~399 and Abell~401 we extract spectra in 3 wide boxes. Two of
them are ($5 \times 15$)~arcmin in size, and the central one is ($5
\times 17$)~arcmin; all three are adjacent to each other and
perpendicular to the line of interest, and they fill the \xmm FOV.

We obtain the data and background spectra, and perform the fitting
procedures in {\sc xspec} as in the previous spectral analysis,
fitting single temperature {\sc mekal} models. Figure ~\ref{T_A399401}
shows the temperature of the gas along the line that intercepts the
two cluster centres, and the resulting temperatures are plotted with
open pentagons. The same figure shows the temperature variations found
in Section~3.2.3 and Section~5: with filled triangles we plot the
temperatures from the sector-N and -S in Abell~401; open triangles
mark the temperatures obtained from the analysis of the 401
observation in Section~5, to the North of Abell~401; filled boxes show
the results for Abell~399 obtained in sector-N and -S.

In order to check the confidence of our results for the temperature of
the gas in the intercluster region, we increase the width of the
previous source regions, so that and we extract the spectra from only
two wider boxes. These results are marked with filled pentagons in the
same figure.

Recently, Loken et al. (2002) derived a `universal temperature
profile' using numerically simulated clusters. They have found that
the temperature declines with the distance ($r$) from the cluster
centre as $1.3 (1 + r/\alpha_x)^{- \delta}$~keV, where $\alpha_x \sim
r_{\rm vir}$. Fits to their simulated data led them to a value for the
exponent of $\delta$ of 1.6, and the normalization of 1.3. When
compared to the data, this profile is in good agreement with De Grandi
\& Molendi (2002), and Markevitch et al. (1998), mainly at large
radii. Loken et al. (2002) found that they could not reproduce the
central core seen in the temperature profiles of De Grandi \& Molendi
(2002). In Fig.~\ref{T_A399401} we show with solid lines the
temperature profile for both of our clusters, as predicted by the
above equation. As virial radii we use the values of $r_{200}$ from Table~5.

It is apparent from Fig.~\ref{T_A399401}, that the temperature of the
intercluster region is enhanced compared to the theoretical
expectations, and the \xmm data to the North of Abell~401.  However,
there is no evidence for a strong shock which could have increased the
temperature to values as high as (15-20)~keV, as seen in the numerical
simulations (e.g., Takizawa 1999).

\section{Abell~401 at large radii}

The region North of Abell~401 was also observed by \xmm in a different
pointing (observation 401). This observation provides us with the
opportunity to investigate the temperature structure of this massive
cluster, at large radii and away from the region that in principle,
should be more disrupted due to the presence of Abell~399. Using
observation 401 we extract the spectra in two adjacent boxes that lie
along the direction that intercepts the two cluster centres. Both
source regions are (5 $\times$ 15)~arcmin in size, and the spectral
analysis is performed in {\sc xspec} as before. We plot the resulting
temperature values in Fig.~\ref{T_A399401} with open triangles. Here
we concentrate on the temperature structure North of Abell~401 (at
negative $R$ in Fig.~\ref{T_A399401}). In this figure we also plot the
temperature profile from the 301 observations, using the derived
temperatures in the sector-N found in Section~3.2.3.

Numerical simulations of structure formation and evolution have
demonstrated that the temperature of present-day clusters declines
with distance from the cluster centre. Theory predicts that this
decline should be significant even within the virial radius (e.g.,
Frenk et al. 1999). This cluster property is encountered in the north
part of Abell~401, as Fig.~\ref{T_A399401} shows. The temperature drop
was initially noted in the analysis of the temperature profiles in the
sectors of Section~3.2.3. The analysis of Section~3.2.3 was restricted
to the inner 10~arcmin, but now we find that the decline carries
further out to $\sim$0.7$r_{200}$. Figure~\ref{T_A399401} shows that
at a distance of $\sim$0.35$r_{200}$ the temperature drops to
approximately (35-60)~percent of its core value. This lower
temperature to the north of the cluster centre can also be seen in
Fig.~\ref{HR}. 

The numerical simulations of Frenk et al. (1999) predict a somehow
milder drop. If we also compare our result with the theoretically
derived universal temperature profile of Loken et al. (2002), we
find once again that our data indicate a temperature decline to the
North of Abell~401 marginally steeper than our
expectations. Although, it has to be noted that the universal
temperature profile of Loken et al. (2002) seems to overpredict the
temperature of the core of Abell~401 (see Fig.~11).

Observationally, if we compare with the temperature profiles derived
from the {\it BeppoSAX} data (De Grandi \& Molendi 2002), and {\it
ASCA} satellite (e.g., Markevitch et al. 1999), we see that the
temperature at large radii to the North of Abell~401 is consistent
with the lower temperature bounds at similar radii.  Recent \xmm
observations of the relaxed cluster Abell~1413 have seen a similar
temperature drop (Pratt \& Arnaud 2002), but again somewhat milder
than the one we observe here. They find that at a similar radius
($\sim$0.35$r_{200}$) the temperature is $\sim$(75-81)~percent its
core value, however in the case of Abell~1413 it is difficult to
define a `core' temperature, since it hosts a mild cooling flow
appearing as a dip in the temperature profile.

It has also to be noted here, that the X-ray flux in the same sector
at large radii from the cluster centre appears somehow suppressed. The
comparisons with the other four surface brightness profiles of
Fig.~\ref{A401T_prof_pies}, and with the global $\beta$-model reveal a
lack of substantial emission at radii $>$6~arcmin along sector-N. This
behaviour can be also seen in Fig.~\ref{cuts}. Additionally, the
1-dimensional analysis of Section~3.1.2 finds the steepest $\beta$-index
($\beta$=0.727; see Table~3) in this sector.  Thus, in summary, the
\xmm data confirm that the temperature to the North of Abell~401
declines. However, this drop seems to be steeper than the theory
predicts and observations of relaxed clusters find. Additionally, it
is accompanied by a drop in the surface brightness distribution in the
same area.

\section{Radio Emission}

Abell~401 was one of the first clusters along with Coma that were
found to host extended radio emission (`radio halo'), that is not
associated with any cluster galaxy (Harris \& Romanishin 1974). Radio
halos are rare radio sources, and they have been found in the inner,
$\sim$1~Mpc of X-ray bright and hot clusters [see Giovannini \&
Feretti (2000) for some recent examples].  They locate the site of
relativistic electrons and magnetic fields in clusters.  Since the
first discoveries, significant advances have been made in
understanding their origin. It has been found, for example, that they
are always associated with clusters that are undergoing disturbances
from recent or on-going merging events (e.g. Buote 2001), although not
all merger remnants host a radio halo. Recent deep X-ray and radio
observations have confirmed that this is true in some cases studied
(e.g., Markevitch \& Vikhlinin 2001).

The most favoured explanation for their origin currently, is that the
electrons have been recently accelerated to the required relativistic
velocities by shock waves or turbulent gas motions, that are generated
during cluster mergers (e.g., Feretti 1999).

\begin{figure*}
\begin{center} 
\leavevmode 
\epsfxsize 0.45\hsize
\epsffile{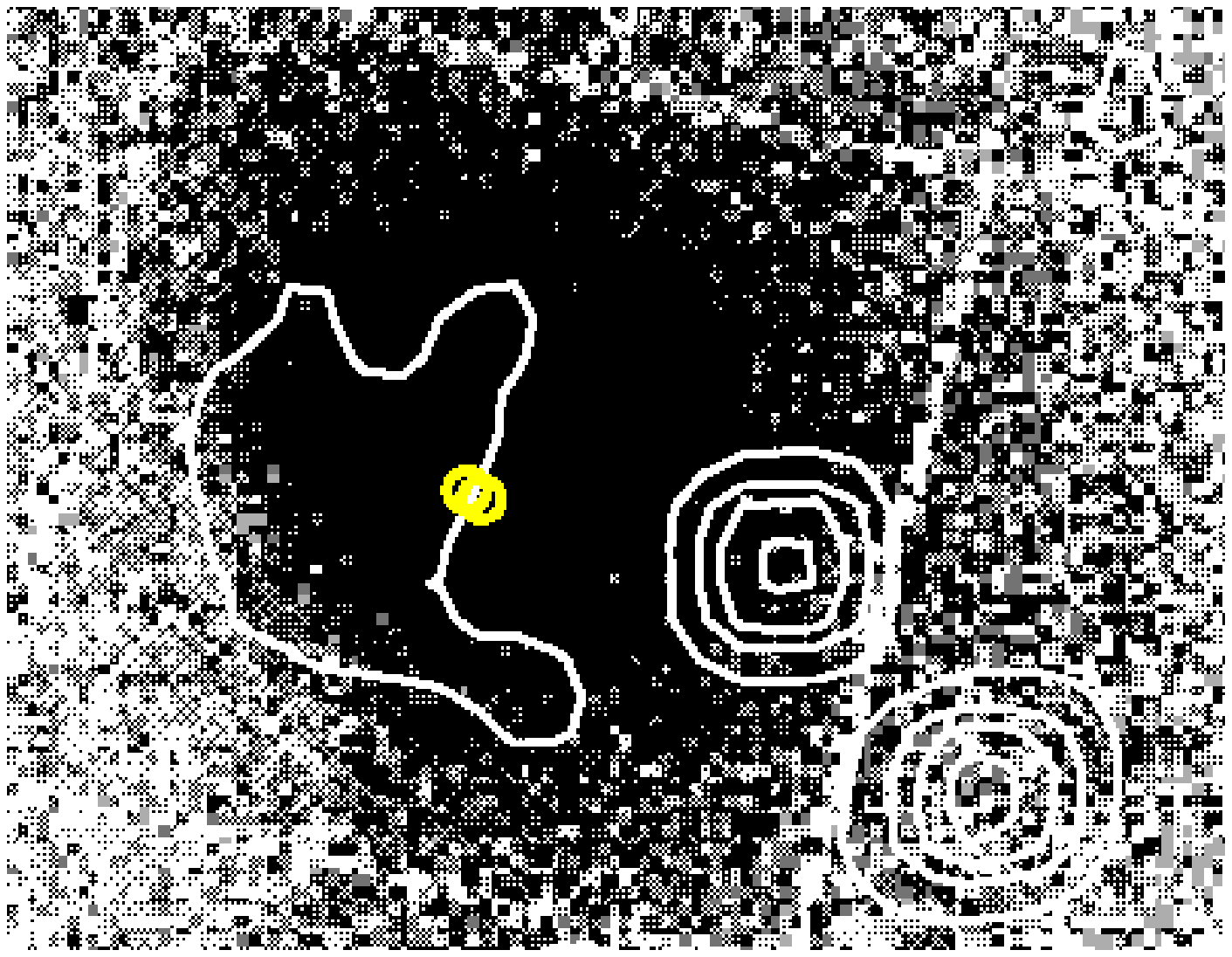}
\leavevmode 
\epsfxsize 0.41\hsize
\epsffile{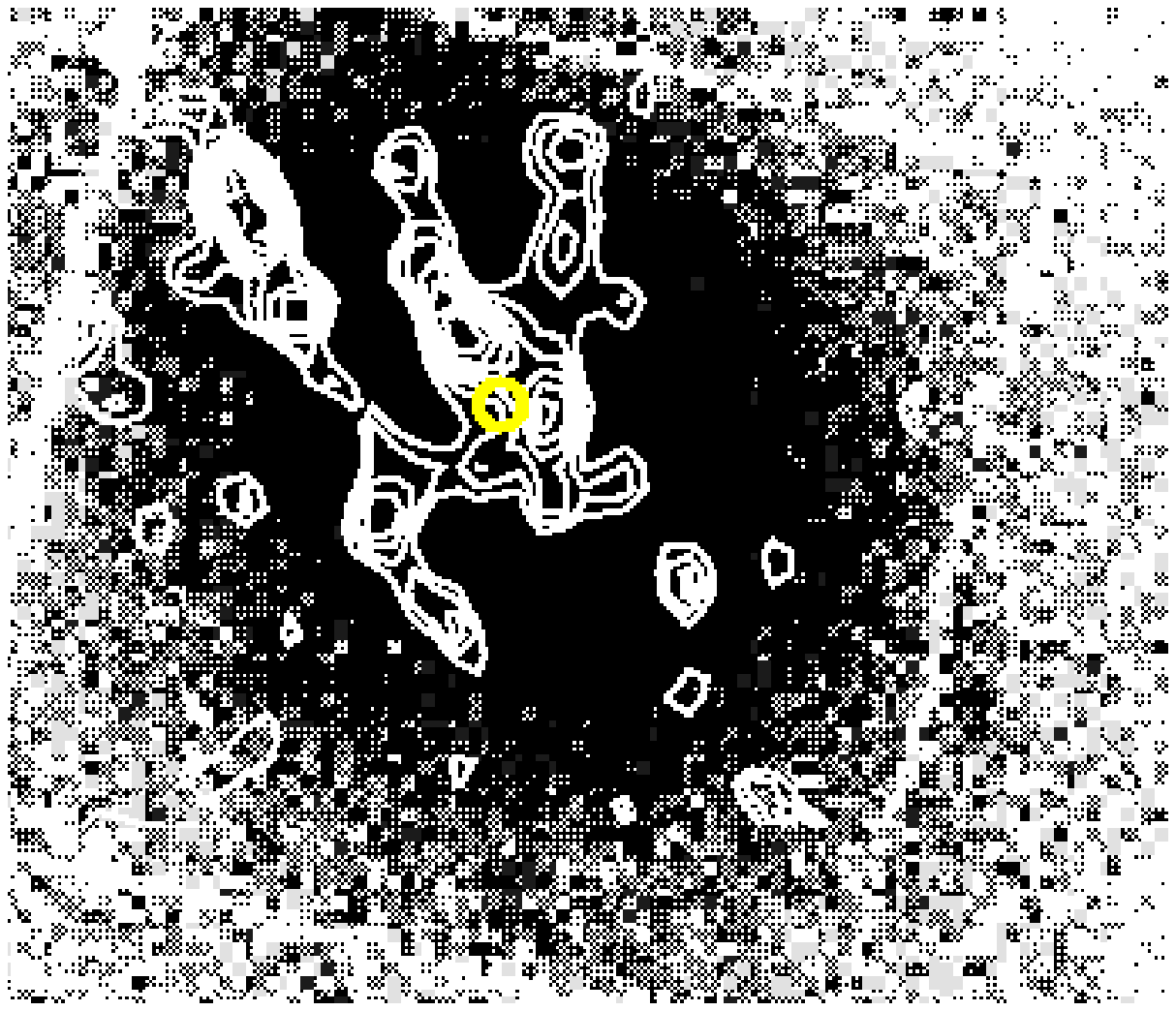}
\end{center}
\caption{The raw \xmm images overlaid by the radio contours. Only data from the two MOS cameras are co-added. Each \xmm image is a circle of R=10~arcmin, and the pixel size is 8~arcsec. The radio contour levels are as in Fig.~\ref{HR}. The location of the central cD galaxies is also marked.}\label{halo}
\end{figure*}

In Fig.~\ref{halo} we overlay the radio contours taken from the NRAO
VLA Sky Survey (NVSS) on the \xmm co-added raw images from the two MOS
cameras. An overlay on the hardness ratio images is shown in
Fig.~\ref{HR}.

The presence of a radio halo in the core of Abell~401 has been
recently confirmed by deep VLA observations (Bacchi et al. 2003). The
same is not true for the halo in Abell~399, where the detection in the
NVSS data is only significant at the 2$\sigma$ level (G. Giovannini,
private communication). This diffuse radio emission was classified as
a radio halo by Giovannini, Tordi \& Feretti (1999) in their search
for new radio halos in the NVSS data, but because the emission is weak
it did not meet their selection criteria for its inclusion in their
final sample, and was not considered further in their publication.

Inspection of the hardness ratio maps of Fig.~\ref{HR} and temperature
profiles of Fig.~\ref{A401T_prof_pies} does not reveal any harder
emission that might be closely associated with the radio halo in
Abell~401, although some small scale variations might be present. On
the other hand, there is more striking evidence that the halo in
Abell~399 might be due to acceleration of electrons by shock waves. It
appears to be associated with the bright edge seen in the raw X-ray
images of Fig.~\ref{halo}. At its location a higher gas temperature
was derived by the fits to the \xmm spectra with thermal models
(Fig.~\ref{A399T_prof_pies}), indicating that the emission is harder
(as also seen in the hardness ratio map of Fig.~\ref{HR}). 

In principle, radio halos might serve as chronometers of mergers. The
radio emission should fade away quickly ($\sim$0.1~Gyr, Jaffe 1977),
hence the presence of a halo suggests that some recent disturbance has
taken place in a cluster.  For example, in Abell~399 we find the radio
halo associated with harder X-ray emission and the edge. Thus, we
might be witnessing there electrons that are currently being
accelerated. On the other hand, in Abell~401 we don't find any close
spatial correlation between the radio halo and the X-ray
emission. This could be explained if turbulent motion after the merger
has given rise to the radio emission.

The peripheries of both clusters also host a number of tailed
radio galaxies (e.g., Burns \& Ulmer 1980; O'Dea \& Owen 1985; Mack et
al. 1993). It has been argued (Bliton et al. 1998), that the existence
of tailed radio galaxies in clusters is closely connected to the
disturbances induced by recent mergers: more tailed sources are found
in clusters that have suffered a recent merger event.

Detailed discussion on the radio properties of the clusters,
their relation to the X-ray emission, and presentation of our recently
approved deep VLA observation is left to a later publication.

\section{The large scale environment of the system}
 
\begin{figure}
\begin{center} 
\leavevmode 
\epsfxsize 0.9\hsize
\epsffile{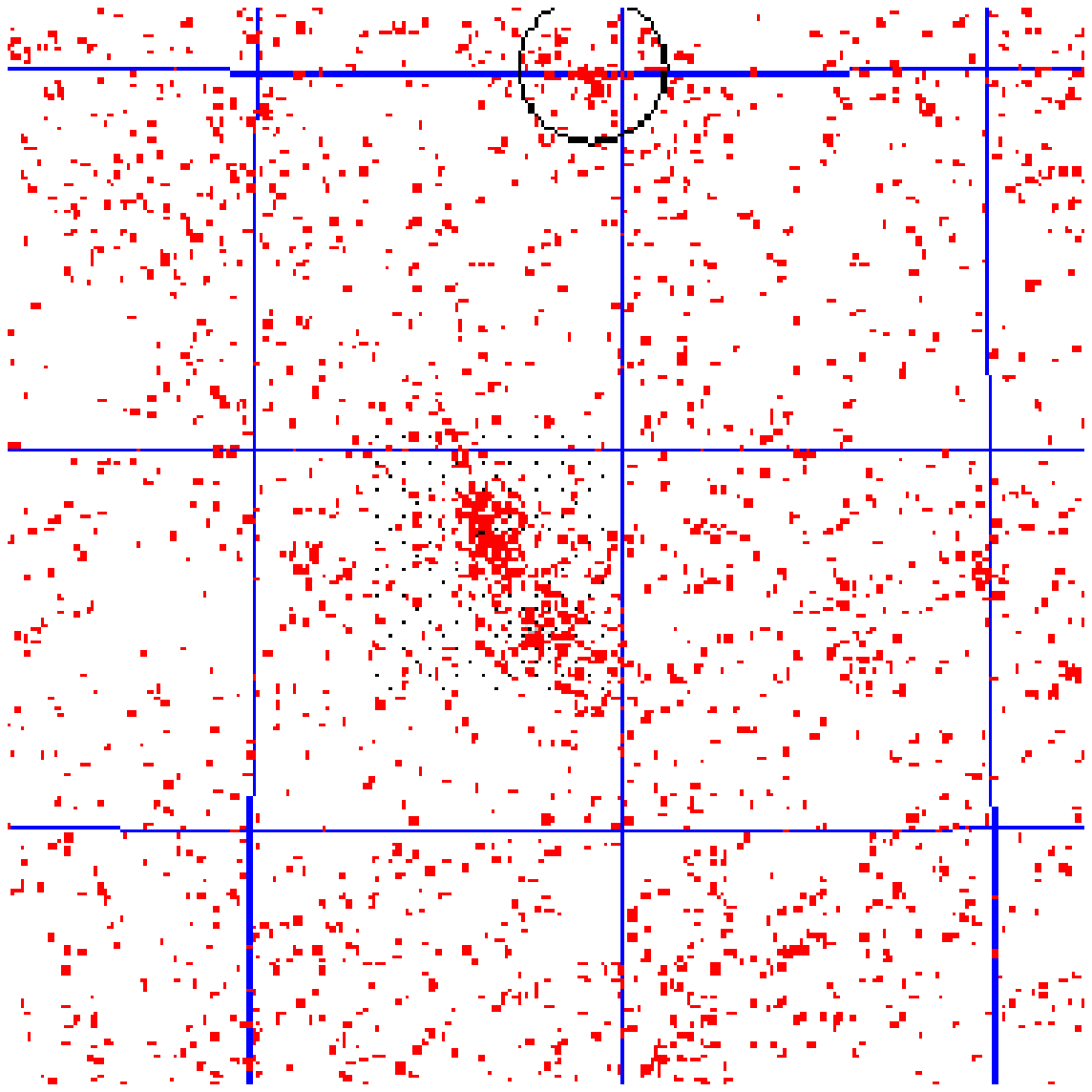}
\caption{The location of all the galaxies in a $(8 \times 8)$~degrees
area around the A399/401 system. No redshift restrictions are
imposed. The concentration around 02h56m58.9s +15d57m02s is Abell~397
at a redshift of z=0.033, and around 03h10m36.1s +09d48m19s is
Abell~421, which is a background cluster, without redshift
information. The location of both clusters is marked by
circles.}\label{galaxies}
\end{center} 
\end{figure}

As mentioned in the introduction, both clusters are at the same
redshift (table~\ref{target_info}), and their projected separation is
only $\sim$3~Mpc. A comparison of this separation with the average
separation of Abell clusters of galaxies ($\sim$20~Mpc), indicates
that Abell~399/401 is an exceptionally close pair of clusters. Such
close separations are normally encountered only in dense
superclusters. It is worth also noting that in Abell cluster
catalogues there are only 46 pairs of clusters out of 786 at a
redshift less than 0.1, that have closer separation than the
Abell~399/401 pair (S. Raychaudhury, private communication). However,
there is no previous record of possible membership of Abell~399 and
Abell~401 in a dense supercluster.

In order to check if the environment of the system is especially
dense, we plot in Fig.~\ref{galaxies} the location of all the galaxies
found in NED in an area of ($8 \times 8$)~degrees around the system. A
redshift restriction between redshifts of 0.06 and 0.09 reveals that
the most of the galaxies recorded in NED come from the study of
Oegerle \& Hill (1993), which was concentrated in a small region
around the cluster pair. In Fig.~\ref{galaxies}, one could see some
smaller concentrations of galaxies around the binary system. Apart
from the foreground Abell~397 (at $\alpha$=03h10m36.1s
$\delta$=+09d48m19s), and the background Abell~421
($\alpha$=03h10m36.1s $\delta$=+09d48m19s), the rest either don't have
redshift information or are background objects.

{\it ROSAT} All-Sky images (RASS) are no more revealing. The analysis
of the RASS images does not show any bright extended source which
might be another cluster close by. Using the RASS images to identify
smaller galaxy groups we search for traces for a
filament/supercluster. We find only 2-3 faint clumps around the pair,
but there is no information recorded about them in any database, and
they don't appear to have bright optical counterparts, forcing us to
conclude that they are background quasars.  Thus, currently, there is
no evidence for a dense supercluster, or filamentary structure around
the cluster pair.

\section{Discussion}

\subsection{Evidence for interactions between the two?}

\subsubsection{Tidal interactions}

\begin{figure}\label{tides}
\begin{center} 
\leavevmode 
\epsfxsize 0.9\hsize
\epsffile{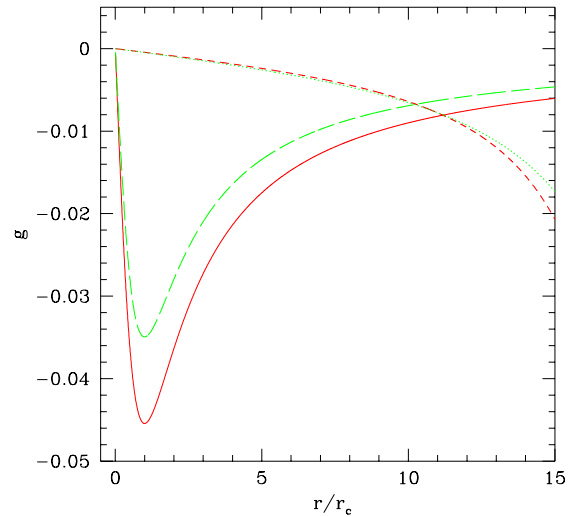}
\caption{Comparison of the gravitational field of each cluster
(A401=solid line; A399=long-dashed line) with the tidal field due to
the presence of the other cluster. The tidal acceleration for A401 due
to the presence of A399 is shown with the short-dashed line; for
A399 with the dotted line. The $x$-axis is the distance from the cluster
centre considered in units of each cluster's core radius (see
Table~5.). The acceleration $g$ is shown in arbitrary
units.}\label{tides}
\end{center}
\end{figure}

The above \xmm analysis finds the cores of both clusters showing
evidence for deviations from the traditional expectations for relaxed
clusters. Given their close proximity, one would think that these
disruptions may be due to tidal interactions. Next, we compare the
strength of the tidal and gravitational fields, and assess the
significance of tidal forces in disrupting the clusters at small and
large radii.

The gravitational acceleration in a cluster can be derived from its
X-ray properties assuming isothermality, as follows:
\begin{equation}
g_{G}(r) \sim kT \frac{\nabla \rho(r)}{\rho(r)}
\end{equation}
where the density distribution [$\rho(r)$] follows the $\beta$-model:
\begin{equation}
\rho(r) = \rho(0) \left[ 1+ \left( \frac{r}{r_{\rm c}} \right)^{2}
\right]^{\frac{-3 \beta}{2}}.
\end{equation}
The central density is $\rho(0) = \mu \ m_{\rm p} \ n_{0}$, with $\mu$
and $m_{\rm p}$ being the mean molecular weight, and proton mass
respectively. 

The tidal field [$g_{T,A}(R)$] experienced by cluster $A$ because of
the presence of cluster $B$ at distance $d$, is :
\begin{equation}
g_{T,A}(R) \sim  g_{G,B}(d-R) - g_{G,B}(d+R), 
\end{equation}
where $R$ is measured from the centre of cluster A. Assuming that both
clusters are spherically symmetric, the accelerations
$g_{G,B}$ can be calculated from eq.~(1).

In Fig.~\ref{tides} we also show the $g_{T}$ for Abell~401 (short
dash) and Abell~399 (dotted line).  This figure demonstrates that the
tides in both cluster cores are negligible compared to the binding
force. The equality $g_{G}=g_{T}$ is reached at a distance of $\sim10
r_{\rm c} \simeq 20$~arcmin. We thus, don't expect that the tidal
field is responsible for the asymmetries we see inside the core
regions of both clusters (the temperature structure, the 2-dimensional
residuals, the presence of the radio halos).

However, tides may become important at larger radii. For example
Fig.~\ref{A399T_prof_pies} and Fig.~\ref{A401T_prof_pies} show that
the surface brightness distribution in Abell~399 appears more extended
in sectors-N and -W.  The same behaviour is observed in Abell~401,
where sector-S (the one that is facing Abell~399) has more extended
distribution than the other three. According to Fig.~\ref{tides} at a
distance of $\sim$8~arcmin $\simeq$4$r_{\rm c}$ the tidal acceleration
is $\sim$10~percent of the gravitational acceleration binding each
cluster, and the difference $g_{G} - g_{T}$ decreases rapidly further
out. Thus, we expect that the tidal field might contribute in
modifying the gas distribution of both clusters at large radii.

Tidal forces will tend to stretch the gas along the line that connects
the two clusters and compress it along the perpendicular
direction. However, for Abell~399/401 the lateral compression
should be rather small compared to the expansion. We find for example,
that for a shell of gas at a distance of (6-7)~arcmin from the cluster
centres the force responsible for the lateral compression is always
$\sim$1 order of magnitude smaller than the force responsible for the
extensions towards the other member of the system.  If the gas
stretches adiabatically we do not expect that its temperature
increases as is evident in Fig.~\ref{T_A399401}. An adiabatic
expansion results in a temperature decrease, which is not what we
observe.

The above discussion demonstrates that tidal forces alone cannot
explain the irregularities seen in the surface brightness profiles at
(8-10)~arcmin radius from the clusters centres. The tidal fields are
only 10~percent of the gravitational field that such radii. Thus,
tidal forces (stretching or lateral compression) have difficulty in
explaining the increased temperature we find in the intercluster gas
in Fig.~\ref{A401T_prof_pies}.

\subsubsection{Compression}

A comparison of the temperature profile to the North of Abell~401 with
the one to the South (see Fig.~\ref{T_A399401}) reveals that the
temperature at (10-20)~arcmin from the centre of Abell~401 to the
South is somehow increased. Although we have already argued that the
temperature drop to the North is steep, a comparison with the
universal cluster temperature profile of Loken et al. (2002) also
indicates that the temperature appears to be elevated.

In the same region, we found in Section~4.1 an increased flux above a
superposition of the two clusters. Thus, in this region we might be
witnessing a compression wave, which could be a sign that the two
clusters are approaching each other, and their gases have already
started interacting.

From the excess of the flux we measured in Section~4.1, we find that
$\Sigma_1/\Sigma_0\simeq 2$ (we use index $1$ for the post compression
quantities, and $0$ are the undisrupted values, and $\Sigma$ is the
photon flux). Using $\Sigma \sim n^{2}$, the compressed region is
denser by $n_1/n_0 \simeq \sqrt{2}$. If the compression is isentropic,
we find that the temperature should have increased by a factor
$T_1/T_0 \simeq 1.7$. Taking the pre-compression temperature ($T_0$)
of this region to be somewhere between the observed value and the
theoretical value of Loken et al. (2002) at the same distance to the
North of the cluster centre (ie., k$T_0\sim4.5$~keV), we find that a
compression could boost the temperature up to $\sim$(7-8)~keV,
somewhat lower than the observed of (7.5-10.0)~keV, but similar within
the errors. Such a compression would reproduce the excess seen in the
surface brightness from this region.

Another possible explanation for the increased temperature and
flux in this region is that it is due to a shock wave. If this is the
case, using eqn.~(2) from Markevitch, Sarazin \& Vikhlinin (1999), we
find that the Mach number of their relative motion should be $M \sim 1.9$,
and the compression factor 1/$x \sim 2.175$. For the above
calculation we use $\gamma = 5/3$, a preshock temperature of
k$T_0\sim4.5$~keV, and postshock temperature of
k$T_1\sim8.75$~keV. Such a shock wave would have increased the flux by
an amount of $\Sigma_1/\Sigma_0= 2.175^{2} \sim 4.73$. Thus, in
Section~4.1 we should have measured a flux of $0.62 \times 10^{-3} \
{\rm cnt \ s^{-1} \ pix^{-1}}$, 2.5 times more than what we actually
observe (see Section~4.1).  

Therefore, we conclude that it is possible that we are witnessing a
compression region to the South of Abell~401, in the direction of
Abell~399, as a shock would have boosted the flux to much higher
levels than the observed. This finding suggests the possibility that
Abell~399 and Abell~401 have already started interacting.  If it is
correct, they should be at early stages of merging, since the \xmm
results are not as dramatic as seen in the simulated images of merging
clusters (e.g., Takizawa 1999) at later stages of their evolution, and
their projected separation is as large as $\sim$3~Mpc.  Note also that
the effects of such an early interaction should not affect the inner
regions of the clusters.

\subsection{Collisions}

\subsubsection{Have they been through each other?}

As mentioned in the introduction, it has been suggested in the past
that the two clusters have already passed through each other (Fabian
et al. 1997). Clusters of galaxies comprise two main components: the
collisionless dark matter, and the collisional gas. When two clusters
collide, the dark matter particles do not interact. Each dark matter
distribution can initially retain its `identity', and after the
collision carry on moving largely undisturbed. On the other hand, the
two gaseous components collide, with dramatic consequences. Shock
waves are formed and propagate within the gas, which is heated to high
temperatures. As a result the gas does not follow the dark matter, but
is left behind, between the two dark matter components. This is
clearly seen in numerical simulations of merging clusters. For
example, Roettiger, Loken \& Burns (1997) show that after core
crossing, the dark matter components carry on moving away from each
each for a time, before they re-collapse to form a common dark matter
halo. However, the two gas components never detach, and don't follow
the two separate dark matter potentials. Hence it seem very unlikely
that a past head-on merger could provide an acceptable model for
Abell~399/401.

\subsubsection{Off-centre collision}

Another possible scenario that could potentially explain the core
disruptions and the large-scale irregularities is that of an
off-centre collision. Abell~399 could have come from the North, and
Abell~401 the South, they have swung closely around each other once in
the past, and have now separated following this close encounter.  A
favourable combination of the closest separation and relative velocity
could disrupt the clusters' central regions, by the propagation of
strong shocks. Details and results of this process can be seen in the
numerical simulations of Ricker (1998) and Ricker \& Sarazin (2001).
Although this explanation for the dynamics of the system might seem
attractive, caution is suggested by the facts that both clusters
appear fairly relaxed at large radii, and that their current
separation is $\sim$3~Mpc.

Based on cosmological arguments, a typical impact parameter for a
cluster collision is $\sim3r_{s}$, where $r_{s}$ is the clusters scale
length [see Sarazin (2001) for further details]. For such an impact
parameter, numerical simulations show that after the first close
encounter, the clusters never acquire a large separation. Tidal
friction brings them together again, after some (3-4)~Gyr, and they
finally merge to form a single remnant. However, larger impact
parameters are possible, so we compare the Abell~399/401 system with
the simulations of Ricker \& Sarazin (2001), for their maximum impact
parameter of $\sim5r_{s}$.  These simulations involve the collision
of a pair of 2~keV clusters, so we must scale their results to the larger
masses of the 7-8~keV Abell~399/401 system. Since $M\propto T^{3/2}$,
our clusters are more massive by a factor of 8. If all lengths are scaled
as $M^{1/3}$, then densities are unchanged (as required for virialised
systems at a given epoch), as is the timescale of the interaction.

The Ricker \& Sarazin (2001) simulations predict that $\sim$2~Gyr
after the first close encounter, the clusters have the largest
separation, which is $\sim1.0$~Mpc. Scaling this distance by a factor
of two (8$^{1/3}$), we expect a maximum separation after a large 
impact-parameter encounter, of no more than 2~Mpc, which is smaller 
than the minimum projected separation of 3~Mpc we observe.
Moreover, the temperature map of the
collision at t=2~Gyr, shows significant structure, with prominent cold
tails emanating from both cluster cores. The \xmm results do not agree
well with this picture. 

In summary, whilst an off-set collision has the potential to explain
both core heating and large scale asymmetries in the gas, the large
separation of Abell~399 and Abell~401, and the modest nature of their
temperature structure, do not appear to be compatible with such a
model.

\subsection{Dynamical state of Abell~399}

Most of the features uncovered by the analysis of the \xmm data lead
us to conclude that Abell~399 is far from being a relaxed cluster, but
it is more likely that we are witnessing a young merger remnant: there
is an edge to the east of the core (sector-E in the previous
discussions), apparent in the \xmm images; the temperature inside this
edge is $\sim$6.7~keV, while immediately outside it increases to
$\sim$8.3 (see Fig.~\ref{A399T_prof_pies}); the image analysis (1- and
2-dimensional) gives uncommonly low values for $\beta$; there is no
evidence for a cooling flow; and there appears to be a weak radio
halo, which seems to coincide with the location of the edge mentioned
above; there are narrow tailed radio galaxies (NATs) at large
distances from the cluster centre. 

Thus, the most plausible explanation for all these properties is that
a small cloud of material has recently fallen towards the centre of
Abell~399, roughly in the East-West direction. The impact has created the
`indentation' (or edge) to the West of the cluster core. Numerical
simulations of merging clusters (e.g., Takizawa 1999; Roettiger, Loken
\& Burns 1997) show that during the infall of even low mass systems
into larger ones such an indentation is created in the gas density
distribution, as the velocity field drags gas towards the cluster
centre of the more massive cluster.  Simulations also predict the
existence of two shock waves that propagate in opposite directions
along the collision axis. We see some evidence for such structures 
in the temperature
profiles obtained in different sectors of Fig.~\ref{A399T_prof_pies}(a):
the `leading' shock (in-front of the infalling blob or group), which
reaches a maximum temperature of $\sim10.2$~keV in the
central $\sim$2~arcmin, appears in sectors-W and -S; the `back' shock
is expected behind the infalling group, and may be seen in the
sector-E at a temperature of $\sim11.5$~keV.  As
expected, the `back' shock is at larger distances than the leading
shock from the cluster centre, as it travels in less dense medium,
going down the density gradient. 

The exact orientation and temperature of the shocks depends of course
on the choice of the sectors, and the 3-dimensional orientation of the
collision direction. However, we believe that the merger is
taking place close to the plane of the sky. In support of this, the analysis of Girardi et al. (1997) finds no substructure in the redshift data of Abell~399, and
the clarity of the `indentation' also suggests that we view the collision 
from an essentially perpendicular perspective.
 
As seen in the simulations of colliding clusters, at the early stages of the event the surface brightness distribution appears elongated along an axis perpendicular to the collision direction. This could explain the orientation of the residuals of the 2-dimensional fits, that appear elongated along the North-South direction.

The last piece of evidence that argues for the merger hypothesis is
the presence of diffuse radio emission in the central regions of the
cluster (see Fig.~\ref{halo} and Fig.~\ref{HR}). Although its presence
is still to be confirmed by deep VLA observations, the \xmm data
provide a first evidence that it might be associated with a shock that
was created during a recent merger event. Thus, we conclude that
Abell~399 is undergoing a merger event with a smaller group, and they
are close to the stage when their two cores collide.

\subsection{Dynamical state of Abell~401}

Abell~401 appears also to be disturbed by a merger: there is 
temperature structure evident in Fig.~\ref{A401T_prof_pies}; the core
region is hotter (see Fig.~\ref{T_prof}); the surface brightness
appears elongated along the North-South direction; there is no cooling
flow; the $\beta$-models that describe best the surface brightness
distribution are not representative of relaxed clusters; bright
residuals are seen in the 2-dimensional spatial analysis; the inner
$\sim$2~arcmin along sector-N and -S are at hotter temperatures that
the rest; there is a significant temperature decline along sector-N,
which is accompanied by a rapidly declining X-ray flux; there is a
radio halo, brighter than the one in Abell~399 (see Fig.~\ref{halo}),
and many tailed radio galaxies.

Thus, a past merger scenario could explain all the above findings, if
one considers that a small group or cloud of material has fallen into
the main cluster from approximately the North, North-West direction,
and now is travelling trough the core region. Its motion has caused
the temperature enhancement in the inner $\sim$2~arcmin along sector-N
and -W. It has also dragged the X-ray emitting material towards the
cluster core, depleting sector-N from its gas, in a similar manner as
in Abell~399 sector-E.
 
\section{Summary and conclusions}

We summarise here the main results of the \xmm study of the
Abell~399/401 binary cluster system and the conclusions about the its
dynamical state that are favoured by them. 
\begin{itemize}
	\item{The \xmm data confirm the lack of cooling flows in the
	cores of both clusters.}  
	\item{The image analysis gives
	$\beta$ values that are lower than the expected canonical
	value of 0.65 found in rich clusters of galaxies. In neither
	cluster is the gas azimuthally symmetric around the central cD
	galaxy.}  
	\item{A 2-dimensional analysis reveals a lop-sided
	excess ($<$200~kpc) around the central galaxies.}  
	\item{There
	is temperature structure in the inner (200-400)~kpc of both
	clusters, which argues for the presence of complex structures
	within the cluster cores, possibly due to shock waves and cold
	fronts.}  
	\item{Abell~399 shows a sharp edge to the East of
	the cluster centre, which is associated with harder X-ray
	emission. The temperature profile to the North of Abell~401
	declines steeper than expected, and there is a lack of flux in
	the same area.}  
	\item{Both clusters appear to host central
	radio halos, although the one seen in the NVSS data of
	Abell~399 requires confirmation.  This halo in Abell~399
	appears to follow an edge apparent in the X-ray images, and is
	associated with harder X-ray emission.}
	\item{There is a plethora of tailed radio galaxies in and
	around both systems.}  
	\item{We find that
	the flux from the space between the two clusters is slightly
	enhanced, above what is expected from the superposition of the
	two clusters. This region appears also hotter than the region
	on the opposite side of the cluster core, and the temperature
	distributions predicted for relaxed clusters by numerical
	simulations.}
	\item{The large scale environment around the binary system appears
surprisingly empty. There is no evidence in the galaxy distribution
and RASS data for the presence of other clusters nearby.}
\end{itemize}

The `substructure' \xmm finds in the core cluster regions cannot be
explained by the tidal field generated by the presence of the other
cluster.  On larger scales, (10-20)~arcmin$\simeq$(0.8-1.6)~Mpc from
the cluster centres, tidal forces become gradually far more important,
in shaping the gas distributions. However, we find that the extensions
of the gases towards the other member of the system seen in the \xmm
data cannot again be solely due to their mutual interactions. If the
gases were extended by gravitational forces, their temperature should
have been lower than what we observe. It is apparent that an
additional driving force is required.

We also find that the properties of Abell~399/401 cannot be reproduced
by scenarios that either involve off-centre cluster collisions, or
that speculate that they have been through each other once. An offset
collision provides the only single mechanism which might explain all
the properties of the system, but that it doesn't seem consistent with
the observed large separation and limited disturbance of the clusters
at large radii from the cluster centres.  On the other hand our
analysis finds evidence for increased flux and temperature in the
region in-between the two. It appears that the two have already stated
interacting mildly, and that we are witnessing a compression region
between them.

The most possible explanation for the properties we find in the core
region of the clusters is that they are due to each ones past merger
activity. It seems that each cluster is a merger remnant. A similar
evolutionary scenario has been recently proposed by Belsole et
al. (2003) to explain the unequal system Abell~1750. Indeed, in a
hierarchical structure formation scenario, binary cluster systems
might be the late stages of cluster formation, just before the final
merger takes place and one rich cluster is formed. The numerical
simulations of Frenk et al. (1996) show two clusters forming at a
close proximity, by the merging of smaller units, before the two
finally merge to form a single rich cluster at present day. 
Additional support for our proposed model comes from the inspection of
simulated X-ray images and temperature maps (from the Simulated X-ray
Cluster Data Archive, http://sca.ncsa.uiuc.edu/). These simulations
show the X-ray flux and temperature stucture of clusters as they are
formed from the continuous accretion of smaller structure. An
extensive search though this archive indicates that pairs of large
clusters are formed, indeed by the merging of filaments from
directions that do not necessarily coinside with the direction to the
nearest massive cluster. Additionally, the development of shock waves
can be witnessed when the two clusters are at close
separations. Additionally, they confirm the presence of features like
the indentaion and shock wave to the East of Abell~399, as being
produced during the infall of a smaller group. Hence, for this pair
of galaxies, we favour a model which combines some early interaction,
with a minor merger history.

\section*{Acknowledgments}

We are grateful to Somak Raychaudhury and Gabriele Giovannini for
helpful discussion and suggestions. We would also like to thank Andrew
Read for the use of his software and advice on \xmm analysis. The
help of Patrick Motl with the Simulated X-ray Cluster Data Archive,
and very useful conversations are also kindly acknowledged. The
referee, Jack Burns, provided us with constructive comments. We would
also like to acknowledge Elena Belsole for useful discussions and
comments. The Digitized Sky Survey, and the NASA/IPAC Extragalactic
Database have been used. The present work is based on observations
obtained with \xmm an ESA science mission with instruments and
contributions directly funded by ESA Member States and the USA (NASA).

\end{document}